\documentclass[11pt]{article}
\pdfoutput=1
\usepackage{amsmath,amsthm,amssymb,amsfonts}
\usepackage[]{natbib} 
\usepackage[pagebackref=false,
            linktoc=page,
            colorlinks=true,
            linkcolor=blue,
            citecolor=blue]{hyperref}
\usepackage{authblk}
\usepackage{pstool,psfrag}
\usepackage{bbm}
\usepackage{bm}
\usepackage{epsfig} 
\usepackage{amsmath}
\DeclareMathOperator*{\argmax}{arg\,max}
\usepackage{makecell}
\usepackage{hyperref}                 
\usepackage{psfrag}
\usepackage[official]{eurosym} 
\usepackage{enumerate}
\usepackage[linesnumbered,lined,boxed,commentsnumbered]{algorithm2e}
\usepackage{pifont}   
\newcommand{\cmark}{\ding{51}}
\newcommand{\xmark}{\ding{55}}
\usepackage{booktabs}
\usepackage{graphicx}
\usepackage{siunitx}  
\usepackage{hyperref}
\PassOptionsToPackage{hyphens}{url}\usepackage{hyperref}
\usepackage{textcomp}
\usepackage{slashbox}
\usepackage{caption}
\usepackage{subcaption}
\usepackage{tabularx}
\usepackage[utf8]{inputenc}
\numberwithin{equation} {section}
\usepackage{graphicx}
\usepackage{mathtools}

\usepackage{geometry}
\usepackage{listings}
\usepackage{color}
\usepackage{fullpage}

\numberwithin{Theorem}{section}
\numberwithin{Definition}{section}
\numberwithin{Lemma}{section}
\numberwithin{Algorithm}{section}
\numberwithin{equation}{section}

\theoremstyle{plain}

\theoremstyle{definition}

\theoremstyle{remark}

\author[$\dagger$\footnote{Corresponding author}]{\textbf{Graeme Auld}}
\author[$\dagger\dagger$]{\textbf{Gabriele C. Hegerl}}
\author[$\dagger$]{\textbf{Ioannis Papastathopoulos}}
\affil[$\dagger$]{\small School of Mathematics and Maxwell Institute,
  University of Edinburgh, Edinburgh, EH9 3FD}
 \affil[$\dagger\dagger$]{\small School of Geosciences, University of Edinburgh, Edinburgh}
\affil[$$]{\small
  G.R.Auld@sms.ed.ac.uk
  $\quad$ gabi.hegerl@ed.ac.uk
  $\quad$ i.papastathopoulos@ed.ac.uk 
}
\date{}
\begin{document}

\title{Changes in the distribution of observed annual maximum temperatures in Europe}
\maketitle

\begin{abstract}
In this study we consider the problem of detecting and quantifying changes in the distribution of the annual maximum daily maximum temperature (TXx) 
in a large gridded data set of European daily temperature during the years 1950-2018.  Several statistical models are considered, each of which models
TXx using a generalized extreme value (GEV) distribution with the GEV parameters varying smoothly over space. 
 In contrast to several previous studies which fit independent GEV models at the grid box level, 
our models pull information from neighbouring grid boxes for more efficient parameter estimation. 
The GEV location and scale parameters are allowed to
vary in time using the log of atmospheric $\textnormal{CO}_2$ as a covariate. 
Changes are detected most strongly in the GEV location parameter with the TXx distributions generally shifting towards hotter temperatures.  Averaged across our spatial domain, the 100-year return level of TXx based on the 2018 climate 
is approximately 2\textdegree{}C hotter than that based on the 1950 climate.  Moreover, also averaging across our spatial domain, the 100-year return level of TXx based on the 
1950 climate corresponds approximately to a 6-year return level in the 2018 climate.
\end{abstract} {\bf Keywords:} Temperature extremes, greenhouse effect,
non-stationary extremes, generalized extreme value distribution, smooth models.

\section{Introduction}  \label{Intro}

The greenhouse effect, whereby increasing levels of greenhouse gases in the Earth's atmosphere leads to a warming of the climate system has been long
understood \citep{charn79} and recent anthropogenic emissions of greenhouse gases are the highest in history \citep{ipcc14}. 
\cite{SR15_Chpt1} estimate that human induced warming in 2017 reached approximately 1\textdegree{}C above pre-industrial levels, and is increasing at a rate of approximately
0.2\textdegree{}C per decade.  \cite{SR15_Chpt3} describe the impacts of 1.5\textdegree{}C global warming above pre-industrial levels on natural and human systems. These impacts include an increase in the frequency and intensity of heavy precipitation events, more frequent marine heatwaves and reduced crop production and yields.

Temperature extremes which may manifest in more intense heatwaves and enhance the risk of fires, pose a risk to human health  \citep[Section 2.3.2]{ipcc14} with the elderly being particularly vulnerable to heat-related mortality \citep{basu02}.  An estimated 40,000-70,000 heat-related deaths were recorded as a result of the summer of 2003 European heatwave \citep{fischer10, robine08} with associated economic losses in excess of \euro{13} billion \citep{bono04}.  Due to the potentially devastating consequences, it is clearly important to understand how the frequency and intensity of temperature extremes may change in a warming climate.

Several previous studies consider changes in the probability distribution of daily temperature and infer that similar changes should also hold for extremes.
\cite{donat12} consider the distribution of daily maximum and minimum temperature on a global scale using observational data and find significant shifts 
in temperature towards higher values in almost all regions but less evidence for changes in variability. Similar conclusions are reported in \cite{weav14} who analyse data from several hundred climate model runs.  \cite{schar04} on the other hand argues that an increase in variability in the daily temperature distribution is required to explain the European heatwave of 2003.

\cite{kiktev03} and \cite{morak13} both find there has been a decrease in the frequency of cold extremes and increase in the frequency of hot extremes, concluding that human-induced forcing has played an important role.  \cite{stott04} consider human influence on the summer heatwave of 2003 and find that ``it is very likely (confidence level $>90$\%) that human influence has at least doubled the risk of a heatwave exceeding this threshold magnitude.''
\cite{zwiers11} use observational data together with climate model output in a detection and attribution study of changes in temperature extremes.  They consider several variables, including annual maximum daily maximum (TXx) and minimum temperatures (TNx), and find evidence for anthropogenic forcing for all variables they consider, with the biggest changes being detected in TNx. More recently, the IPPC report \citep{ipcc_summary21}
concluded that ``human-induced climate change is the main driver'' of the increase in intensity and frequency of hot extremes.

\begin{figure}[h]
\begin{center}
\includegraphics[scale=0.5]{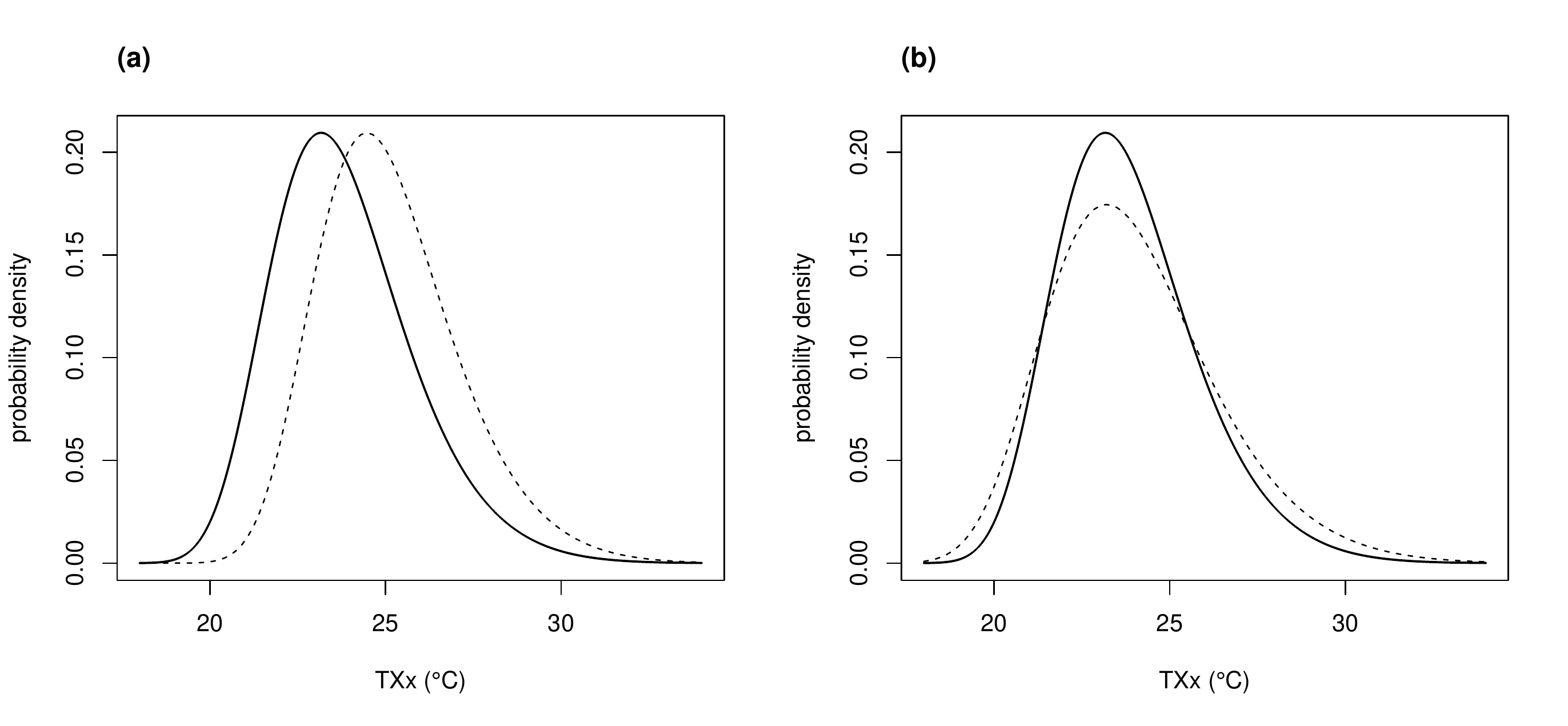}
\caption{\textbf{(a)} The solid black curve shows a hypothetical probability density function of TXx. The dashed curve illustrates the effect of a shift in the 
location of the distribution.  \textbf{(b)} The solid black curve is as in (a) but the dashed curve illustrates a change in the scale of the distribution.}
\label{IntroPlot}
\end{center}
\end{figure}

In this paper we consider statistical models for the variable TXx at approximately 12,000 locations of a gridded data set in a large subset of Europe.
We consider the question of whether, over various large subregions of Europe, there is evidence for changes in the distributions of TXx and if so how are such changes best described.  Our approach can, informally, be viewed as macroscopic, since we are interested in detecting changes in TXx
on a large scale rather than at any one specific geographic location.  
We fit statistical models that allow for changes in both the location and scale of the TXx distributions.  A change in the location of the TXx distribution corresponds to a horizontal shift of the distribution with the mean and all quantiles being shifted by the same amount.
A change in scale corresponds to a horizontal stretching or compression of the distribution, which in turn changes measures of variability, such as the variance of TXx.
Figure \ref{IntroPlot} illustrates both of these effects for a hypothetical TXx distribution.

Most of the studies mentioned above treat the data occurring at different geographic locations in an independent manner, fitting separate statistical models to the data at each location.  One difficulty with this approach in the context of extremes is that, as extreme observations are by definition rare, we will only have 
a small sample at each location, making precise estimation of trends problematic. Although it may be unreasonable to assume a common trend at every geographic location of a large spatial domain,
we would nonetheless expect nearby regions to be similarly affected by climate change.  There are several classes of models, such as varying coefficient models \citep{hastie93}
or geographically weighted regression models \citep{bruns98} that allow us to borrow strength from neighbouring locations to obtain spatially coherent estimates of trends.  Varying coefficient models allow for regression coefficients, e.g., trends, to vary smoothly over a spatial domain, and may be formulated under the generalized additive model (GAM) framework of \cite{wood_book}
and consequently fit with the R \citep{R21} package \texttt{mgcv}. 
As we work with a gridded domain, we consider a discrete analogue of varying coefficient models that are based on Gaussian Markov random fields \citep{rue05} and fall under the general smooth modelling framework of \cite{wood16}. 
Previous studies that make use of GAMs or smooth models for modelling environmental extremes include \cite{demoulin05} and \cite{young19}.


The structure of the paper is as follows.  Section \ref{DataSec} describes the data that is used for fitting statistical models. Section \ref{MethSec} describes the models that are considered and Section \ref{ResultsSec} presents the results which are summarized in Section \ref{DiscSec}.  

\section{Data}  \label{DataSec}

We use the daily E-OBS data, publicly available through the European Climate Assessment and Dataset (ECA \& D) project.  The E-OBS datasets are based on observational data from an underlying network of weather stations interpolated on to a regular $\ang{0.25}\times\ang{0.25}$ grid.  Although the data set covers all of Europe as well as North Africa and the Middle East,  the spatial density of the underlying weather station network that is used to estimate the gridded areal averages is highly variable over the domain.

As explained in \cite{hof12}, when the station density is low, the estimates of the areal averages are over-smoothed which leads to a negative bias of the true areal average. Although this bias applies to the whole distribution, the effect is more pronounced for higher quantiles, so that the weather extremes are disproportionately affected.\, 
This effect is more of a concern for precipitation than temperature due to its discontinuous nature. Similar comments are found in \cite{hof09} who remark that
``overall extreme temperature events will be quite well represented'' although ``E-OBS is fundamentally limited by its underlying station network.''   

Both \cite{hof09} and \cite{hof12} express reservations about using E-OBS for the detection of trends in extremes, mainly due to inhomogeneities that may be present in the underlying station data, i.e., non-climatic factors such as changes in instruments or observing practices, as well as the fact that the network density is not homogenous in time. 
The documentation accompanying the release of E-OBS v18.0 also comes with a similar warning: ``it remains the case that many of the input station series have not been homogenized and at present we caution against the use of E-OBS for evaluating trends'' \citep{Cornes18}.   For this reason we use the E-OBS v19.0e data which is a version of E-OBS that has been homogenized by the ECA \& D in collaboration with the Horizon 2020 EUSTACE Project. 
The method by which the data was homogenized is described in \cite{squintu19}.

\begin{figure}[h]
\begin{center}
\includegraphics[scale=0.45]{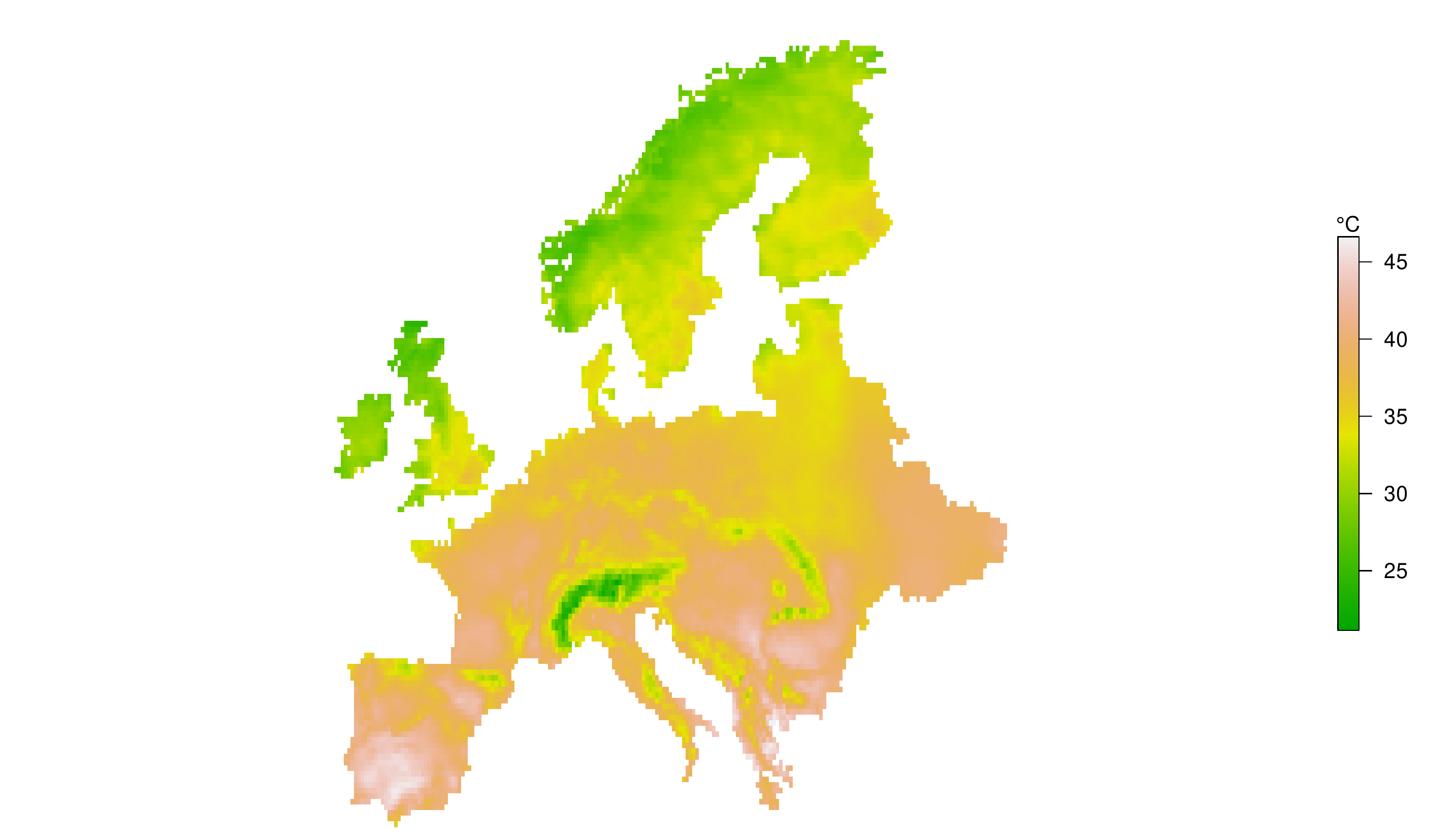}
\caption{The spatial domain considered, showing the maximum value of the variable TXx (annual maximum daily maximum temperature)
at each grid box during the period 1950-2018.}
\label{AbsMaxPlot}
\end{center}
\end{figure}

A plot of the station network density used in E-OBS can be found in \cite{schrier13} which shows that the highest density is in Central Europe and is particularly low in North Africa, the Middle East and Eastern Europe.  The data set covers the years 1950 to 2018.  There is some missing data which is mainly in the early years although there is also little data for Russia in the last ten years.  We consider a large subset of the full domain covered by E-OBS, shown in Figure \ref{AbsMaxPlot}, that
has reasonable station density.  The values displayed in Figure \ref{AbsMaxPlot} are the maximum value of TXx recorded during our study period, 1950-2018, and thus correspond to the absolute maximum observed temperature at each location that we consider. 
With the exception of the United Kingdom and the Republic of Ireland, small islands off the mainland are excluded.
We set the value of TXx at a given location in a given year as missing if there are more than 10 missing observations in that year.  This is a slightly stricter criterion than is typically applied in other studies, e.g., \cite{zwiers11} allow for 15 missing observations.  

For atmospheric $\textnormal{CO}_2$ concentration, we use data from \cite{ssp20}. The historical, observation based data, is only available until the year 2015 after which projections are provided until the year 2500 under different socio-economic scenarios.  For the years 2016 to 2018, we took values from a mid-range scenario, namely SSP2-4.5, which are similar to the values recorded at the Mauna Loa Observatory \citep{keeling76}. 

\section{Methods}  \label{MethSec}

\subsection{Generalized extreme value distribution}  \label{GEV_Sec}

Our approach is based on fitting generalized extreme value (GEV) distributions to the TXx values at each grid box. 
Another possible, and theoretically well-founded approach to modelling extremes is the peaks over threshold method \citep{davismit90}, which models the distribution of exceedances above some large threshold rather than the maximum over large blocks of observations.  We prefer the block maxima approach in our setting due to the difficulty in making a principled choice of appropriate thresholds at such a large number of spatial locations as well as the sensitivity of inference to the choice of thresholds which is exacerbated by the presence of trends \citep{northjon11}.

Just as variations in the mean of a large number of independent and identically distributed random variables are naturally modelled by a normal (Gaussian) random variable, variations in the sample maximum are most naturally modelled by a GEV random variable with distribution function 
\begin{equation}  
G(y; \bm{\Psi}) = 
\text{exp}\bigg[-\bigg\{1 + \xi\bigg(\frac{y-\mu}{\sigma}\bigg)\bigg\}_+^{-1/\xi}\,\bigg],   \label{GEV_CDF}
\end{equation} 
where $\bm{\Psi} = (\mu, \sigma, \xi)$ is a vector of parameters that relate to the location, scale and shape of the distribution respectively and
 $x_+ = \max(x, 0).$  The scale parameter $\sigma$ is strictly positive whereas $\mu$ and $\xi$ may be any real number. The case where $\xi = 0$ in 
(\ref{GEV_CDF}) should be interpreted as the limit as $\xi\to 0$, which gives rise to the distribution function 
$G(y) = \text{exp}[-\text{exp}\{-(y-\mu)/\sigma\}], y\in \mathbb{R}.$  
The formal justification for using the GEV distribution to model TXx comes from the extremal types theorem 
\citep[Theorem 3.1.1]{cole01}.  

The GEV family of distributions encompasses three distinct types of distribution according to the sign of the shape parameter $\xi$.  The cases $\xi > 0$, $\xi=0$ and $\xi < 0$ are known as the Fr\'echet, Gumbel and Weibull classes respectively, which fundamentally differ from each other in the behaviour in their upper (right) tail.
For the Fr\'echet class, the right tail decays according to a power law and for larger values of $\xi$, extremes take on an increasingly volatile nature such as might be expected in financial \citep{resnick07} or hydrological \citep{katz02} applications. The Weibull class has an upper bounded right tail, with $\mu - \sigma/\xi$ the theoretical maximum possible value, whereas the Gumbel class is an intermediate case with a light upper tail that decays exponentially.  Typically, when modelling annual maximum temperatures, we expect to be in either the Weibull or Gumbell class, i.e., $\xi \leq 0.$ 

Suppose that, in grid box $i$ of the E-OBS data, we observe the annual maximum temperature in a total of $n_i$ years, say $t_{i1}, t_{i2},\ldots, t_{in_i}$, which for most grid boxes is each year from 1950 to 2018 inclusive so that $n_i = 69$.  Let $y_{it_j}$ denote the annual maximum temperature in grid box $i$ in year $t_{ij}$, $1\leq j \leq n_i$. 
 If we assume that $y_{it_1}, y_{it_2},\ldots, y_{it_{n_i}}$  are independent realizations of a GEV random variable with distribution 
function (\ref{GEV_CDF}) and parameters $\bm{\Psi}_i = (\mu_i, \sigma_i, \xi_i)$,
then one way to estimate $\bm{\Psi}_i$ is to find the parameter configuration
 $\hat{\bm{\Psi}}_i  = (\hat{\mu}_i, \hat{\sigma}_i, \hat{\xi}_i)$ that maximizes the log-likelihood function $l(\bm{\Psi}_i)$, i.e.,
$ \hat{\bm{\Psi}}_i =  \argmax_{\bm{\Psi}_i}\,l(\bm{\Psi}_i)  $
where $l(\bm{\Psi}_i) = \sum_{j=1}^{n_i}\text{log}\,g(y_{it_j}; \bm{\Psi}_i)$ and $g(y)=\frac{dG}{dy},$ with $G$ as in (\ref{GEV_CDF}), is the GEV density function.
The explicit expression for the log-likelihood function is 
\begin{equation} \label{loglik1}
l(\bm{\Psi}_i) = -\text{log}\,\sigma_i - (1 + \xi_{i}^{-1})\sum_{j=1}^{n_i}\text{log}\bigg[1 + \xi_{i}\bigg(\frac{y_{it_j} - \mu_{i}}{\sigma_{i}} \bigg) \bigg]_+ 
 - \sum_{j=1}^{n_i}\bigg[1 + \xi_{i}\bigg(\frac{y_{it_j}-  \mu_{i}}{\sigma_{i}}\bigg) \bigg]^{-1/\xi_i}_+,
\end{equation}
with the case $\xi_i = 0$ being defined by continuity. 
Although the annual maximum temperatures may not be independent, it is assumed that the dependence between maxima from different years is sufficiently weak that 
the log-likelihood (\ref{loglik1}) may be used as a reasonable approximation of the true likelihood.
The resulting maximum likelihood estimator $\hat{\bm{\Psi}}_i$ is a consistent and asymptotically normal estimator of the true parameter vector provided that $\xi > -1/2$ \citep{smith85, bucher17}. 
\cite{hosk85} gives details for implementing the Newton-Raphson method to find the parameters $\bm{\Psi}$ that maximize (\ref{loglik1}) and several R \citep{R21} packages, e.g., \texttt{ismev} \citep{ismev} or \texttt{extRemes} \citep{extRemes}, provide routines for estimating the GEV parameters using maximum likelihood. 

Having estimated $\bm{\Psi}_i$, we may estimate the temperature $y_p$ that is exceeded in grid box $i$ in a given year 
with probability $p$ by solving the equation $G(y_p) = 1-p$ for $y_p$, with $G$ as in (\ref{GEV_CDF}).  This yields the estimate $\hat{y}_p$ of $y_p$
\begin{equation} \label{RetLev1}
\hat{y}_p = 
\begin{cases}
\hat{\mu}_i - \frac{\hat{\sigma}_i}{\hat{\xi}_i}[1 - \{-\text{log}(1-p)\}^{-\hat{\xi}_i}], \quad & \hat{\xi}_i \neq 0, \\
\hat{\mu}_i - \hat{\sigma}_i\text{log}\{-\text{log}(1-p)\}, \quad  & \hat{\xi}_i = 0.
\end{cases}
\end{equation}
The quantity $y_p$ is known as the return level with associated return period $1/p$.  This definition of return period in terms of the reciprocal of an annual maxima quantile is not universally adopted and other definitions may be found in the literature, see \cite{cooley13} for a discussion. 

From (\ref{RetLev1}) we see that errors in the estimated value of $\xi_i$ may be magnified in the estimate of $y_p$, e.g., due to the dependence of $y_p$ on $1/\xi_i$ 
and the fact that in many environmental applications, $\xi_i$ is close to zero.  When the sample size is small,
the maximum likelihood estimator of $\xi_i$ can have high bias, leading to absurd estimated return levels that are orders of magnitude beyond what would be deemed physically possible, and several authors \citep{coles99, martins00} have proposed adjustments to the log-likelihood function (\ref{loglik1}) to overcome this difficulty. 

\begin{figure}[h]
\begin{center}
\includegraphics[scale=0.6]{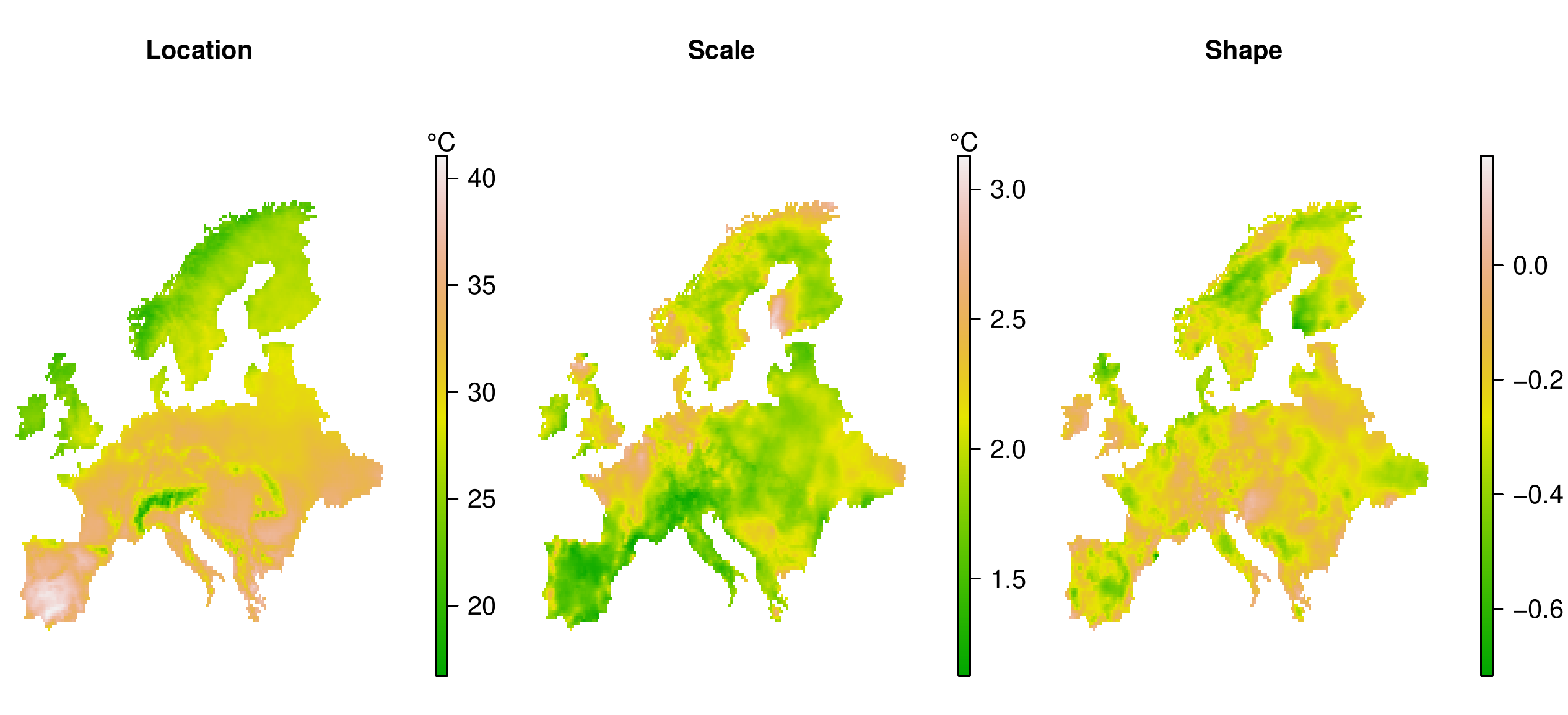}
\caption{Point estimates of the GEV parameters, fitted to the TXx values (\textdegree{}C) separately at each grid box, using the method of L-moments. }
\label{PointEstGEV}
\end{center}
\end{figure}

An alternative to maximum likelihood estimation that is more robust to small sample sizes is the method of L-moments, or equivalently, probability weighted moments \citep{hosk_etal85, hosk90}.   The L-moment estimates of the three GEV parameters at each grid box of the E-OBS data are shown in Figure \ref{PointEstGEV}, which were calculated using the R \citep{R21} package \texttt{extRemes} \citep{extRemes}.
The method of L-moments is often used in a spatial setting as part of a regional frequency analysis (RFA), as set out in \cite{hoskwall_book}. 
In a RFA, data from different regions that are deemed to be sufficiently homogenous is pulled together to increase the sample size, and hence reduce the
uncertainty in parameter estimates. One difficulty with RFA is the sensitivity of the results to the method used to identify homogenous regions.
Moreover, L-moments estimation is not suited to statistical modelling as it does not allow the GEV parameters to depend on the values of covariates 
such as the atmospheric level of $\textnormal{CO}_2$.  Our approach, described in Section \ref{StatMods}, has a similar motivation to RFA but as it is likelihood based, allows for the inclusion of covariates.  

\subsection{Statistical models}  \label{StatMods}

In this section we describe the statistical models that we fit to the E-OBS data. For computational convenience, and also to allow for the possibility that different models may be better suited to different regions, we partition our spatial domain in to eight subregions, which are defined in Table \ref{RegionSplit}.  
The abbreviations used for the regions are meant to be informative and correspond, roughly, to South-Western Europe and France (SWFR), Central and South-Central Europe (CESC), Central Europe 2 (CE2), South-Eastern Europe (SE), Eastern Europe (EAST), Norway and Sweden (NRSW), Finland (FIN) and the United Kingdom and Republic of Ireland (UKRI).

\begin{table}  
\centering
\caption{The various subregions of the domain that the models from Table \ref{ModTrends} are separately fit to.} 
 \label{T1}
\begin{tabular}{|c | c | }
 \hline 
\textbf{Abbreviation}  &  \textbf{Countries included}  \\
\hline
SWFR & Portugal, Spain, Andorra, France \& Monaco \\
\hline
CESC & Germany, Netherlands, Luxembourg, Belgium, \\ 
         &   Italy, Switzerland, Austria \& Denmark      \\
\hline
CE2 & Poland, Czech Republic, Slovakia, Hungary \& Slovenia  \\   
\hline
SE &   Croatia,  Kosovo, Montenegro, Bosnia and Herzegovina,  \\    
     &   Serbia, Macedonia, Albania, Greece \& Moldova  \\
\hline
EAST & Ukraine, Belarus, Lithuania, Latvia \& Estonia \\ 
\hline
NRSW & Norway \& Sweden  \\
\hline
FIN & Finland  \\
\hline
UKRI & United Kingdom \& Republic of Ireland \\
\hline
\end{tabular}
\label{RegionSplit}
\end{table}

For statistical modelling of TXx, the log-likelihood function (\ref{loglik1}) may be considered too simple in at least two respects.  Firstly, it assumes that the GEV parameters at a given grid box remain fixed from year to year, whereas a potentially more realistic model would allow them to change over time.
Secondly, we expect that the parameters of neighbouring grid boxes are more likely to be similar than those of grid boxes that are far apart and (\ref{loglik1}) does not allow us to incorporate this belief.  Moreover, maximizing (\ref{loglik1}) separately for each grid box $i$ may lead to highly uncertain or unrealistic parameter estimates due to the small sample available at each grid box and for the purposes of statistical inference we also run in to problems with multiple comparisons \citep{farco08, chen17}. 

The dependency of the GEV parameters on time can be linked to that of a climatological covariate, and for this purpose we will use the atmospheric concentration of $\textnormal{CO}_2$, which is the dominant greenhouse gas that affects temperature \citep{stips16}. 
More specifically, we will use the derived covariate $x_t = \text{log}\,(\textnormal{CO}_{2,t} / 280)$ where $\textnormal{CO}_{2,t}$ is the 
 atmospheric concentration, in parts per million (ppm), of $\textnormal{CO}_2$ in year $t$ of our study period, $1\leq t \leq 69$, with $t=1$ corresponding to the year 1950, and 280 ppm is, approximately,  the pre-industrial atmospheric concentration of $\textnormal{CO}_2$.
The reason for using the log-transformed covariate $x_t$ rather than the raw $\textnormal{CO}_{2,t}$ values
is due to the approximate logarithmic effect of $\textnormal{CO}_2$ on temperature \citep{jones_hegerl98}. 

We assume that the the annual maximum temperature in grid box $i$ in year 
$t$, $1\leq t \leq 69$, follows a GEV distribution with time varying parameter vector  
$\bm{\Psi}_{it}= (\mu_{it}, \sigma_{it}, \xi_{it})$.  A simple model we may consider, to which we will add further structure and covariates later, is 
\begin{align}
\mu_{it} &= \mu^{(0)}_i + \mu^{(1)}_ix_t \\
\sigma_{it} &= \sigma_i  \\
\xi_{it} &= \xi_i.   
\end{align}
For this model, only the GEV location parameter is time varying. The intercept parameter $\mu^{(0)}_i $ can be interpreted as the value of the GEV location 
parameter in grid box $i$ if atmospheric $\textnormal{CO}_2$ were at its pre-industrial level, whereas the slope parameter $\mu^{(1)}_i$ is the change in the location
parameter that would occur if $x_t$ increased by 1 unit, i.e., if atmospheric $\textnormal{CO}_2$ increased by a factor of $e \approx 2.718$.  Over the course of our study period, atmospheric $\textnormal{CO}_2$ has increased by a 
factor of  1.31, i.e., $\textnormal{CO}_{2,69} = 1.31\text{CO}_{2,1}$. 
As in Section \ref{GEV_Sec}, if in grid box $i$, we have observations in years $t_{i1}, t_{i2},\ldots, t_{in_i}$, 
then writing $\bm{\theta}_i = (\mu^{(0)}_i, \mu^{(1)}_i, \sigma_i, \xi_i)$,  the log-likelihood function for $\bm{\theta}_i$  is 
\begin{align}   \label{GridLogLik}
	l_i(\bm{\theta}_i) = &-n_i\text{log}\sigma_i - (1 + \xi_i^{-1})\sum_{j=1}^{n_i}\text{log}\bigg[1 + \xi_i\bigg(\frac{y_{it_j} - \mu^{(0)}_i - \mu^{(1)}_ix_{t_j} }{\sigma_i} \bigg)   \bigg]_+ \notag \\
& - \sum_{j=1}^{n_i}\bigg[1 + \xi_i\bigg(\frac{y_{it_j}-  \mu^{(0)}_i - \mu^{(1)}_ix_{t_j} }{\sigma_i}\bigg) \bigg]^{-1/\xi_i}_+.
\end{align}
If there are $n$ grid boxes in total, then maximizing (\ref{GridLogLik}) separately for each $i$, $1\leq i \leq n$, is equivalent to jointly maximizing 
the function $l(\bm{\theta}) = \sum_{i=1}^{n}l_i(\bm{\theta}_i)$ where $\bm{\theta} = (\bm{\theta}_1, \bm{\theta}_2, \ldots, \bm{\theta}_n)$ 
is the vector containing the parameters for all grid boxes.  This is because the $j$-th term in the summation defining $l(\bm{\theta}) $ contains only the parameters of 
grid box $j$ which occur at no other terms in the summation and so the maximization problem is separable.   Rather than fitting a model where every grid box is forced to ``learn for itself'', one way to obtain parameter estimates for all grid boxes that are spatially coherent, and reduce uncertainty in the estimates, is to add a term to the objective function $l(\bm{\theta}) $ that 
will penalize model fits where there is too much local variation in the parameters.  As we are working on a discrete gridded domain, it is natural to use a 
penalty that is based on Gaussian Markov random fields \citep{rue05}.

The Gaussian Markov random field (GMRF) penalty allows us to formalise the belief that grid boxes that are near to each other are more likely to have parameter values that are similar than those that are far apart. In order to define the GMRF penalty we are required to specify a neighbourhood structure for our domain. Specifically, for each grid box $i$ we are required to specify the set of neighbours of $i$, which we denote by N$(i),$ and define to be those grid boxes that share a common (grid box) edge with $i$.  Thus, for most grid boxes, N($i$) will consist of four neighbours, but in some cases there may be less than four, e.g., if $i$ lies on the boundary of the domain. 
Now, if we define $\overline{\text{N}(i)}$ to be the set of neighbours $j$ of $i$ with $j > i$, then placing a GMRF penalty on the GEV shape parameters $\bm{\xi} = (\xi_1,\xi_2,\ldots,\xi_n)^T$ for example, amounts to the penalty term
\begin{equation}
P_{\mbox{\tiny GMRF}}(\bm{\xi}) = \sum_{i=1}^{n}\sum_{j\in\overline{\text{N}(i)}}(\xi_i - \xi_j)^2  = \bm{\xi}^T\textbf{S}\,\bm{\xi}
\end{equation}
where the penalty matrix $\bm{S}$ satisfies  $S_{ij} = -1$ if $i\in \text{N}(j)$ and $S_{ii} = n_i$ where $n_i$ is the number of neighbouring regions of $i$, not including $i$. Clearly $P_{\mbox{\tiny GMRF}}(\bm{\xi})$ takes larger values when there is more local variability in the $\xi_i$, $1 \leq i \leq n$. 
If we also impose a GMRF penalty on each of the location intercept, slope, scale and shape parameters, then writing 
$\bm{\mu}_0 = (\mu^{(0)}_1, \mu^{(0)}_2,\ldots, \mu^{(0)}_n)^T, \bm{\mu}_1 = (\mu^{(1)}_1, \mu^{(1)}_2,\ldots, \mu^{(1)}_n)^T,
 \bm{\sigma} = (\sigma_1, \sigma_2,\ldots, \sigma_n)^T$ and $\bm{\xi} = (\xi_1,\xi_2,\ldots, \xi_n)^T $, the objective function that we seek to maximize is the penalized log-likelihood
\begin{equation}  \label{FullObjective}
 \sum_{i=1}^n l_i(\bm{\theta}_i) - (\lambda_1\bm{\mu}_0^T\textbf{S}\,\bm{\mu}_0 + \lambda_2\bm{\mu}_1^T\textbf{S}\,\bm{\mu}_1 + \lambda_3\bm{\sigma}^T\textbf{S}\,\bm{\sigma} + \lambda_4\bm{\xi}^T\textbf{S}\,\bm{\xi})
\end{equation}
with $ l_i(\bm{\theta}_i) $ is as in (\ref{GridLogLik}) and $\lambda_i > 0$, $1\leq i \leq 4$ are constants.  The constants $\lambda_i$, $1\leq i \leq 4,$ are smoothing,
or regularization, parameters that specify the relative priorities given to the competing goals of smoothness and fitting a model that closely matches the observed data.
If, for example, $\lambda_4$ is extremely large, then we would obtain a fit with a low amount of variability in $\bm{\xi}$.   Rather than subjectively choosing a value for the smoothing parameters, they may be selected in a more objective manner, by marginal likelihood maximization as in
\cite{wood16}, and this is the approach taken in R \citep{R21} package \texttt{evgam} \citep{evgam} that we use for model fitting. 

The objective function in (\ref{FullObjective}) is the same as the log posterior obtained from a Bayesian model specification where
the observations from grid boxes $i$ and $j$, $i\neq j$, are conditionally independent given $(\bm{\theta}_i, \bm{\theta}_j),$ and 
independent intrinsic Gaussian Markov random field priors \citep[Chapter 3]{rue05} are placed on each of the parameter vectors $\bm{\mu}_0, \bm{\mu}_1, \bm{\sigma}$ and $\bm{\xi}$.   The conditional independence assumption is standard in Bayesian spatial \citep{banerjee04} and latent Gaussian \citep{rue09} modelling, but note that this 
is not the same as assuming that the observations from grid boxes $i$ and $j$,  $i\neq j$, are independent. 
The fact that the smoothing parameters, which correspond to hyperparameters in a Bayesian analysis, are found by maximizing 
a marginal likelihood means that the fitting approach may be regarded as empirical Bayes. The fitted value, $\hat{\bm{\theta}}$, of $\bm{\theta}$ that maximizes (\ref{FullObjective}) is then the mode of the posterior distribution, also known as the maximum a posteriori probability (MAP) estimate.\, 
Credible intervals for any component of $\bm{\theta}$, or linear combination of components, can be computed based on the asymptotic normality of 
$\hat{\bm{\theta}}$ \citep[Section 2]{wood16}.

The model that has been described so far in this section, contains only a single covariate in the GEV location parameter, and we have seen how the effect of this covariate on the annual maximum temperature can be modelled as smoothly varying over space by using the GMRF penalty.  The value, $x_t$, of the covariate in year $t$ is taken to be the same at each grid box in year $t$, so that the covariate is spatially homogenous.  We may also include spatially varying covariates in our model, and for this purpose we will include elevation (km) as a covariate in the GEV location parameter.  From Figure \ref{PointEstGEV}, it is clear that larger values of elevation tend to be associated with smaller values of the GEV location parameter.  The framework of \cite{wood16} allows us to model covariates as having a general smooth, rather than simply linear, effect on the location parameter.  
However, based on exploratory model fits we find it adequate to specify elevation as having a linear effect on the GEV location term.  
We also consider models that have trends in the GEV scale parameter.  In total, we consider five different models that differ from each other with regards to the inclusion of the covariate $x_t = \text{log}(\textnormal{CO}_{2,t}/ 280)$.   
For each model, it is assumed that $Y_{it} \sim \text{GEV}(\mu_{it}, \sigma_{it}, \xi_{it})$ where, as before, $Y_{it}$ corresponds to the value of TXx in grid box $i$ in year $t$.  The differences between the models with regards the inclusion of trends are summarized in Table \ref{ModTrends}. 

The most complex model is Mod4 which corresponds to the following formulas for the GEV parameters
\begin{align*}
 \mu_{it}  &= \mu^{(0)}_i + \mu^{(1)}_ix_t +\beta\,elevation_i,   \\ 
 \text{log}\,\sigma_{it}  &= \sigma^{(0)}_i + \sigma^{(1)}_i x_t,   \\
 \xi_{it}  &= \xi_{it},
\end{align*}
where $x_t = \text{log}(\text{CO}2_t/280)$,  $elevation_i$ corresponds to the elevation (km) of grid box $i$ minus the mean elevation across all grid boxes, and independent GMRF penalties are placed on $\bm{\mu}_0 , \bm{\mu}_1, \bm{\sigma}_0 = (\sigma^{(0)}_1, \sigma^{(0)}_2,\ldots, \sigma^{(0)}_n)^T , \bm{\sigma}_1 = (\sigma^{(1)}_1, \sigma^{(1)}_2,\ldots, \sigma^{(1)}_n)^T$  and $\bm{\xi}.$
The fixed effect $\beta$ gives the change in the GEV location parameter for a one km increase in elevation.
The trend in scale parameter is modelled using the log link to ensure that the scale remains positive. Mod1, Mod2 and Mod3 are each special, simpler, 
cases of Mod4. In particular, Mod1 has $\bm{\mu}_1 = \bm{0}$ and $\bm{\sigma}_1 = \bm{0},$ corresponding to the situation where there is no climate change signal detectable in TXx, whereas Mod2 and Mod3 correspond to the cases $\bm{\sigma}_1 = \bm{0}$ and $\bm{\mu}_1 = \bm{0}$ respectively.  
Mod5 has only a trend in the GEV location parameter but this is modelled as a fixed effect, i.e., the same trend is assumed at each geographic location,
corresponding to $\mu_{it} =  \mu^{(0)}_i + \mu_1x_t + \beta\,elevation_i,$ and note that the trend $\mu_1$ does not depend on $i$.

\begin{figure}
  \centering
\begin{subfigure}{0.9\textwidth}
  \centering
\includegraphics[scale=0.5]{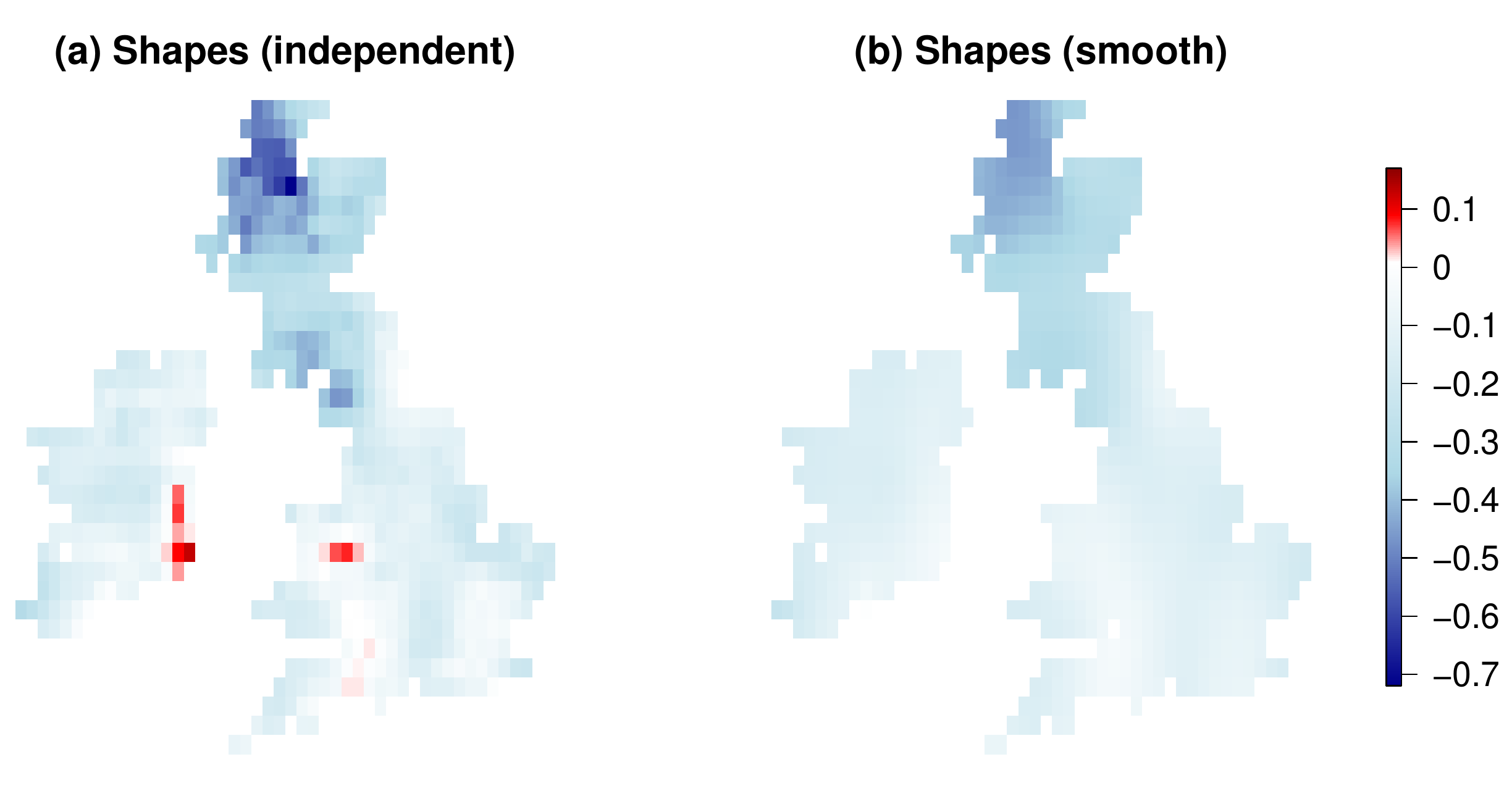}   
  \label{PRmean}
\end{subfigure}

\begin{subfigure}{0.9\textwidth}
  \centering
\includegraphics[scale=0.5]{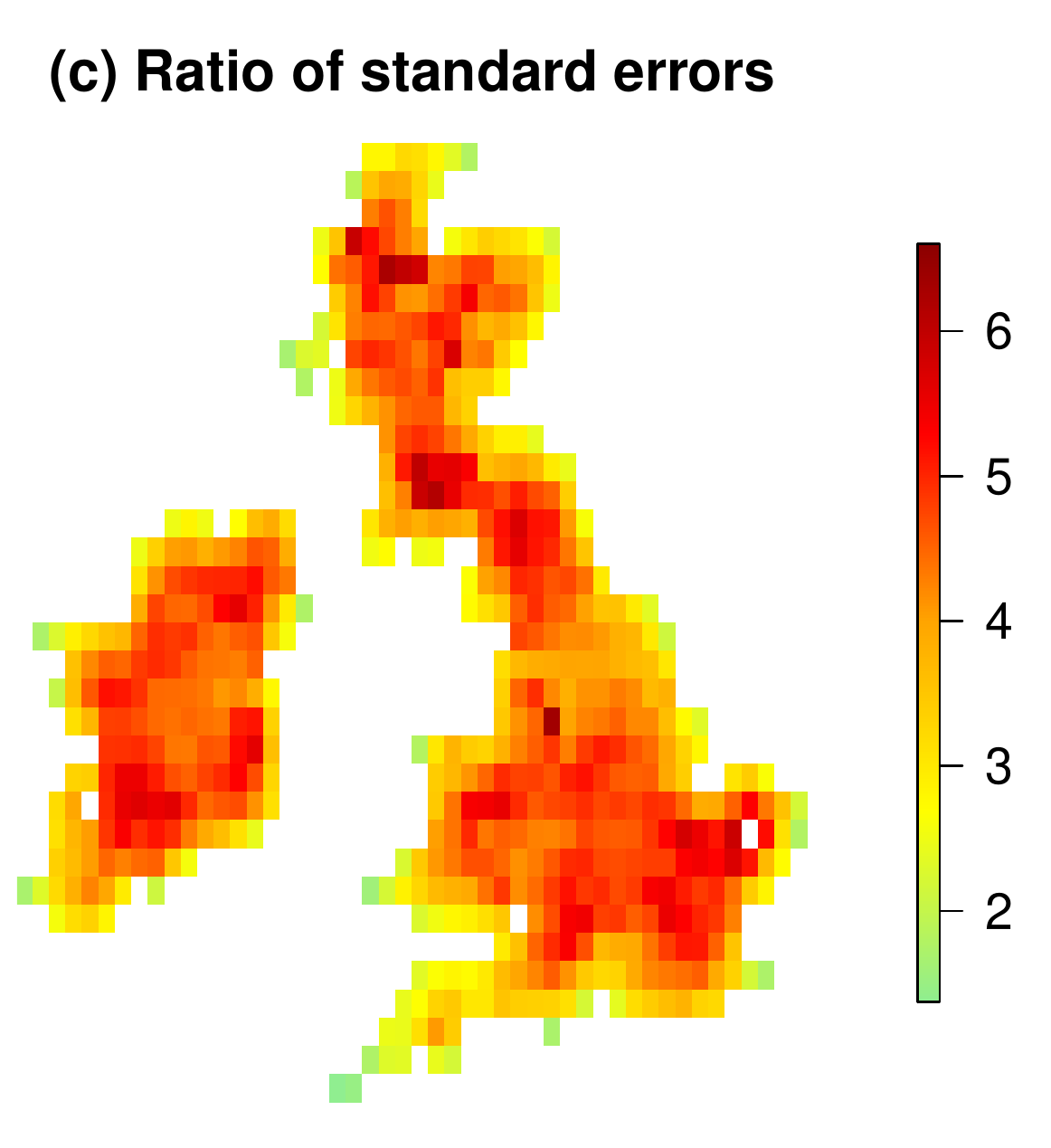} 
  \label{PRsd}
\end{subfigure}
\caption{Plots (a) and (b) show the fitted GEV shape parameters, $\xi$, for independent, i.e., separate, grid box fits compared to a smooth model fit using the GMRF penalty respectively. Plot (c) shows the ratio of the parameter uncertainty, as measured by the standard error, for the independent grid box fits relative to the smooth model fit which uses information from neighbouring grid boxes. }
\label{IndepVersusSmooth}
\end{figure}

To illustrate the effect and benefit of using the GMRF penalty, we compare, for region UKRI, the independent grid box fits based on maximizing 
(\ref{GridLogLik}) separately for each $i$, using R package \texttt{ismev}, with joint maximization of the penalized log-likelihood (\ref{FullObjective}) for the smooth model using \texttt{evgam}. 
Figure \ref{IndepVersusSmooth} (a) and (b) show the fitted values of the GEV shape parameters, $\bm{\xi}$, for the independent grid box and smooth model fits respectively.  Although the broad spatial pattern of fitted shapes is similar in both cases, the fitted shapes for the smooth model encompass the more plausible range of -0.47 to -0.001 compared to the independent grid box fits which range from the extremely short tail of -0.71 to the heavy tailed case of 0.16. The reduction in uncertainty that occurs by including neighbouring information in the model fitting procedure is also illustrated in 
Figure \ref{IndepVersusSmooth} (c).  This shows the ratio in parameter uncertainty, as measured by the standard error, for the independent grid box model fits relative to the smooth model.  For the independent grid box model fits, the standard errors were computed based on asymptotic normality of the maximum likelihood estimators, and for the smooth model, standard errors were also computed based on asymptotic normality using the results of 
\cite{wood16} Section 2.  All the ratios are greater than one indicating that we have reduced the parameter uncertainty at all grid boxes with the mean ratio 
being equal to approximately 4, which represents the average reduction in uncertainty achieved by including neighbouring information in the model fitting. 


\begin{table}  
\centering
\caption{Comparison of Mod1-Mod5 according to the inclusion of a trend in $x_t = \text{log}(\text{CO}2_t/280)$ in the GEV
location ($\mu$) and log scale (log\,$\sigma$) parameters and whether these trends are assumed to vary over space (spatially varying) or 
the same trend is assumed at each grid box (spatially homogenous).  All models include an altitude covariate. } 
 \label{T1}
\begin{tabular}{|c | c | c | c |}
 \hline 
\textbf{Model}  & Trend in $\mu$  & Trend in log\,$\sigma$ \\
\hline
Mod1 &  \xmark &  \xmark  \\
\hline
Mod2 &  \cmark   &  \xmark \\
         & (spatially varying)    &  \\
\hline
Mod3 & \xmark  &   \cmark  \\   
         &            & (spatially varying) \\
\hline
Mod4 &  \cmark  &   \cmark  \\    
         & (spatially varying)    & (spatially varying)  \\
\hline
Mod5 &  \cmark  & \xmark  \\
         &  (spatially homogenous) & \\
\hline
\end{tabular}
\label{ModTrends}
\end{table}

\subsection{Changes in return levels and risk ratios} \label{Sec_RL_RR}

In Section \ref{GEV_Sec} we defined the return level with return period $1/p$ to be the temperature that, in a stationary climate, is exceeded in a given year with probability $p$.  In a non-stationary climate, this quantity will typically vary from year to year.  For grid box $i$ in year $t$, $1\leq t \leq 69$, we modify (\ref{RetLev1}), and define
$y_{it}(p)$ by
\begin{equation} \label{RetLev2}
y_{it}(p)= 
\begin{cases}
\mu_{it} - \frac{\sigma_{it}}{\xi_{it}}[1 - \{-\text{log}(1-p)\}^{-\xi_{it}}], \quad & \xi_{it} \neq 0, \\
\mu_{it} - \sigma_{it}\text{log}\{-\text{log}(1-p)\}, \quad  & \xi_{it} = 0.
\end{cases}
\end{equation}
The quantity $y_{it}(p)$ may be interpreted as the return level with return period $1/p$ if the climate were stationary in the same state as in year $t$.  
We consider, for each grid box $i$ the difference $y_{i69}(0.01) - y_{i1}(0.01)$, which tells us the difference in the 100-year return levels in grid box $i$ based on
the 2018 and 1950 climates.  We quantify the uncertainty in the return level differences via simulation.  In particular, at each grid box, we simulate 2000 values of $\mu_{it}, \sigma_{it}$ and $\xi_{it}$ for $t=1$ and $t=69$ from their posterior distributions 
and calculate the corresponding differences $y_{i69}(0.01) - y_{i1}(0.01)$.  We then calculate the 0.025 and 0.975 quantiles of these 2000 values to give us approximate $95\%$ credible interval for the return level differences. 

Another way that we quantify changes in the distribution of the annual maximum temperatures is via risk ratios \citep{NatAcad}. 
If $Y_{it}$ denotes a random variable having the same distribution as the annual maximum temperature in grid box $i$ in year $t$, i.e., GEV with parameter vector 
$\bm{\Psi}_{it}= (\mu_{it}, \sigma_{it}, \xi_{it})$, we consider the risk ratio
\begin{equation} \label{RR_Defn}
\frac{\mathbb{P}\{Y_{i69} > y_{i1}(0.01)\}}{\mathbb{P}\{Y_{i1} > y_{i1}(0.01)\}}  =
\frac{\mathbb{P}\{Y_{i69} > y_{i1}(0.01)\}}{0.01}. 
\end{equation}
The value of the ratio (\ref{RR_Defn}) then tells us, in grid box $i$, how many times more likely the 100-year return level based on the 1950 climate is to be exceeded in the 2018 climate. 
We quantify the uncertainty in the estimated risk ratio in the same way as the return level difference, via simulation. 

\section{Results}  \label{ResultsSec}

All models were fit using R \citep{R21} package \texttt{evgam} \citep{evgam} on a Dell PowerEdge R430 computer running Scientific Linux 7 with four 
Intel Xeon E5-2680 v3 processors. As this is a shared departmental cluster, our access was restricted to 10 cores. 
For a given fixed subregion in Table \ref{RegionSplit}, we assess the performance of each of the five models in Table \ref{ModTrends} 
using the Akaike information criterion (AIC) as well as using several scoring rules \citep{gneit07} in a 5-fold cross-validation \citep{stone74, hastie09}.
The scoring rules considered are the squared error (SE),
Dawid-Sebastiani (DS), continuous ranked probability (CRP) and the weighted continuous ranked probability (WCRP) scores.  
All the scoring rules we consider are negatively oriented so that a smaller score indicates better performance.
Further information on the scoring rules and the cross-validation procedure can be found in Appendices \ref{ModScoring} and \ref{CV_Sec} 
respectively.
For a given subregion in Table \ref{RegionSplit} the cross-validation was performed in parallel using the R package \texttt{parallel}. The total compute time for the full cross-validation on all subregions was between 2 and 3 weeks. 

Table \ref{ModScores} shows the mean scores of each model by region from 5-fold cross-validation. For all regions, Mod4, which includes a spatially varying trend in both the GEV location and log-scale parameters, is the best performing model, closely followed by Mod2 which only contains the spatially varying trend in the GEV location parameter. 
Mod5 which has a fixed effect of the covariate $x_t = \text{log}\,(\textnormal{CO}_{2,t} / 280)$ in the location parameter, i.e., a constant trend at all grid boxes in a given subregion, is generally the next best performing model. Mod1 which has all of  the GEV parameters fixed in time is the worst performing model in all regions according to all scores, followed by Mod3 which has only a spatially varying trend in the GEV log-scale parameter.

Table \ref{PVals} shows the p-values, by region, of the hypothesis test that the CRP scores for Mod4 and Mod2 (the two best performing models) are pairwise exchangeable. The testing procedure is described in Appendix \ref{CV_Sec}.  Failure to reject the null hypothesis would, informally, mean that the performances of Mod4 and Mod2 are statistically indistinguishable.  
In all regions the null hypothesis of pairwise exchangeability is rejected at the 0.05 and 0.01 levels of significance, after adjusting the raw p-values in Table \ref{PVals} using the Bonferroni correction for multiple comparisons,  giving evidence in favour of the better performing Mod4.  
We also perform the same test for the WCRP scores of Mod4 and Mod2. 
For all regions the null hypothesis of pairwise exchangeability of WCRP scores is rejected at the 0.05 and 0.01 level of significance giving evidence in favour 
of Mod4.  

\begin{figure}
\begin{center}
 \includegraphics[scale=0.6]{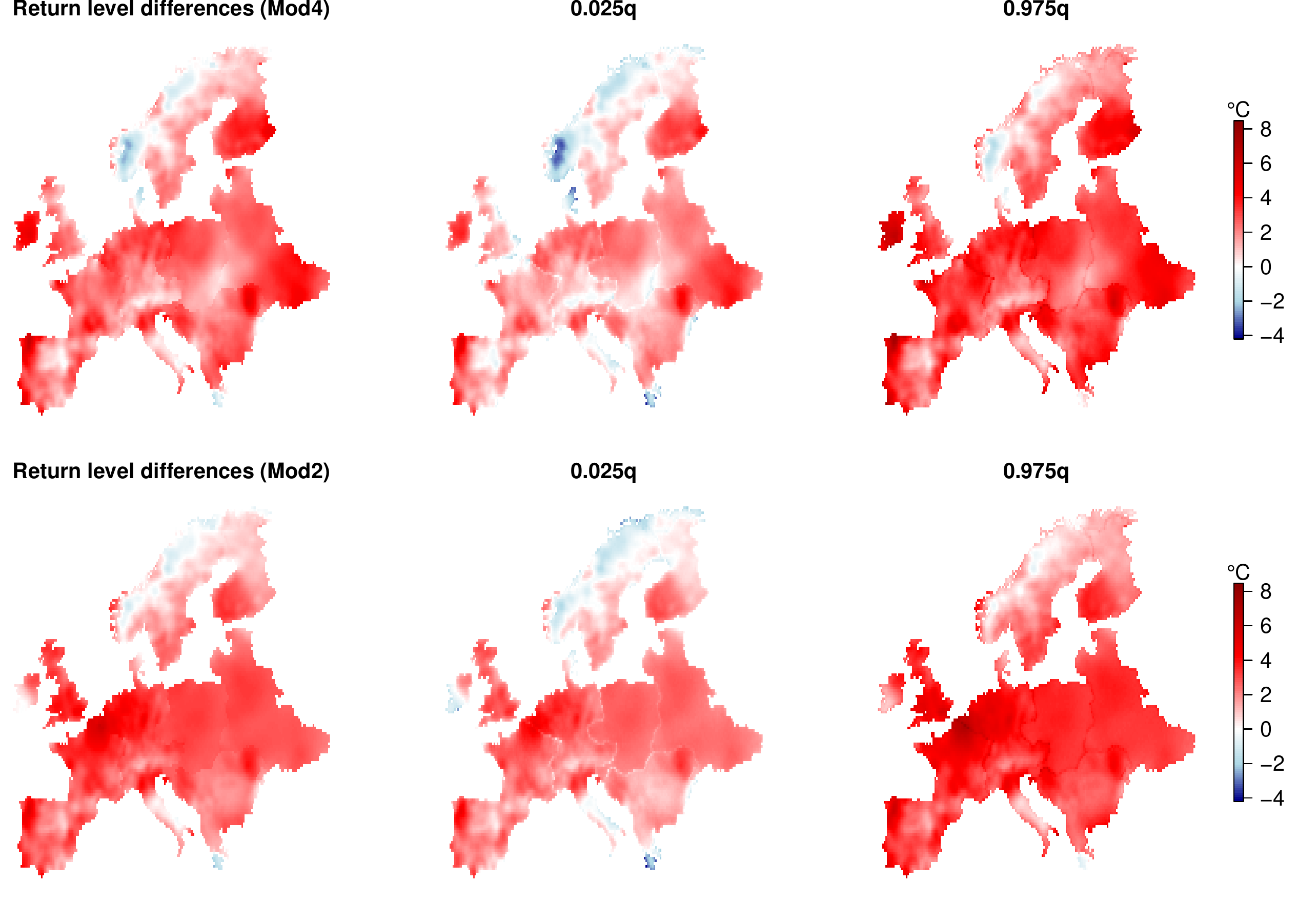}  
\caption{The difference, in \textdegree{}C, in the 100 year return levels based on 2018 and 1950 climates (2018 return level subtract 1950 return level) and approximate 95\% credible interval limits, calculated by Monte Carlo simulation as described in Section \ref{Sec_RL_RR}, for Mod4 (top row) and Mod2 (bottom row).  Mod4 includes a trend in the GEV location and log-scale parameters using the covariate log($\textnormal{CO}_{2,t}/280$), whereas Mod2 only includes a trend in the GEV location parameter.}
 \label{RLDiffRas}
\end{center}
\end{figure}

\begin{figure}
\begin{center}
  \includegraphics[scale=0.6]{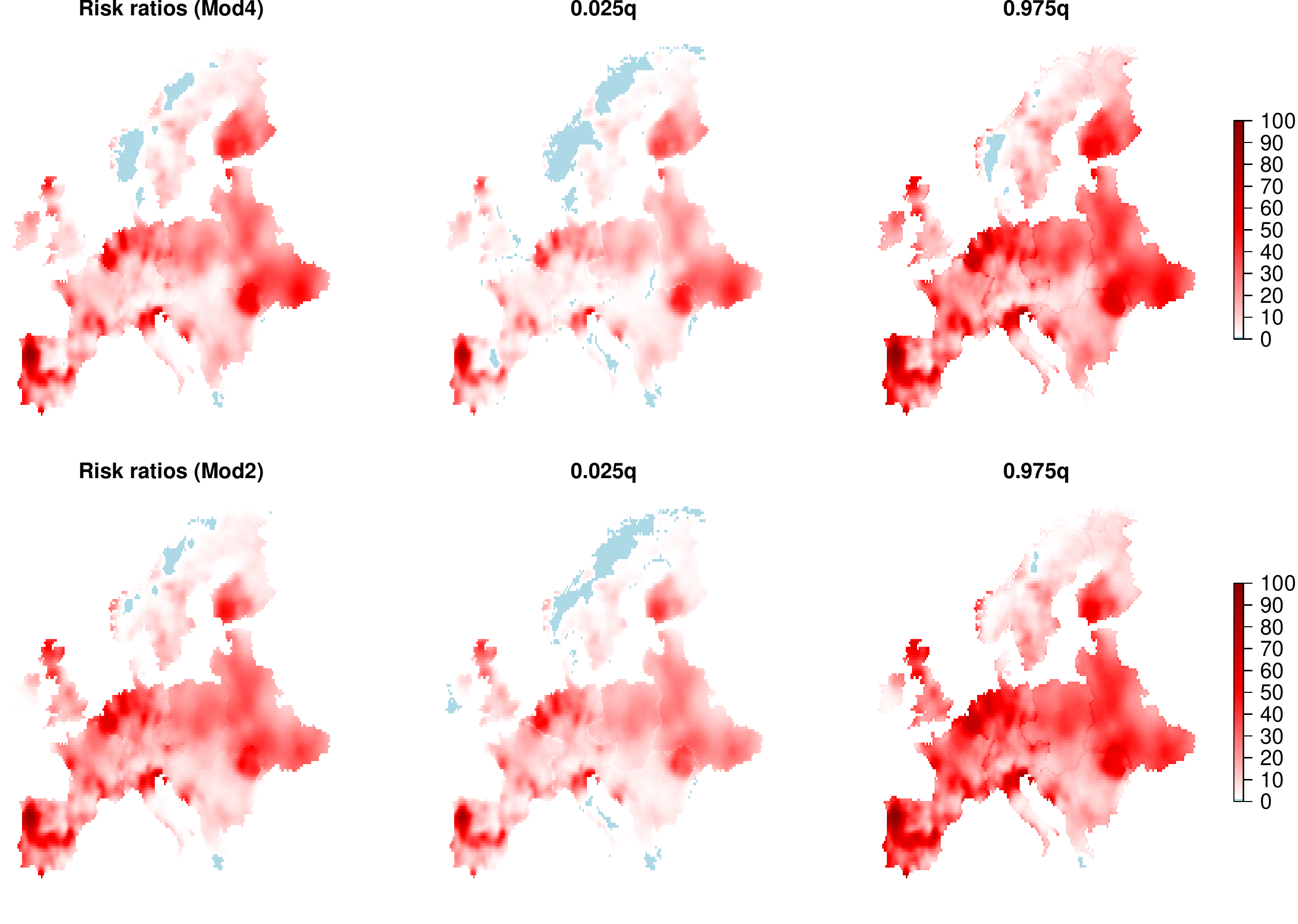}  
\caption{Risk ratios and approximate 95\% credible interval limits, calculated by Monte Carlo simulation as described in Section \ref{Sec_RL_RR}, for Mod4 (top row) and Mod2 (bottom row). Light blue corresponds to a risk ratio of less than one.
Mod4 includes a trend in the GEV location and log-scale parameters using the covariate log($\textnormal{CO}_{2,t}/280$), whereas Mod2 only includes a trend in the GEV location parameter.}
  \label{RiskRatios}
\end{center}
\end{figure}

Figure \ref{RLDiffRas} shows the difference in 100-year return-levels based on 2018 and 1950 climates, for both Mod4 and Mod2, along with approximate 95\% credible intervals, calculated using Monte Carlo simulation, as described in Section \ref{Sec_RL_RR}. The corresponding risk ratio plots are shown in Figure \ref{RiskRatios}. Mod2, which only has a trend in the GEV location parameter,
tends to find slightly larger increases in the 100-year return levels return levels based on the 2018 climate compared to 1950, 
but the spatial pattern broadly agrees with that produced by Mod4, and similar comments apply for the risk ratios. 
All regions show, on average, significant increases in return levels and risk ratios greater than one, as shown in Table \ref{MeanDiff}. 
Region NRSW, comprising Norway and Sweden, stands out as having relatively moderate changes compared to the other regions, with the mean increase over the region, 
in the 100-year return level being around 0.6\textdegree{}C.  There are several locations within this region where negative changes are detected although upon inspecting the
approximate 95\% credible intervals in Figures \ref{RLDiffRas} and \ref{RiskRatios}, most of these changes do not appear significant.
Even in this region of relatively modest changes, the 100-year return level based on the 1950 climate is estimated to be,
on average, between 4 and 5 times more likely to be exceeded in the 2018 climate. 
The region EAST, comprising Eastern European countries shows the most dramatic increases,
with the mean 100-year return level difference of around 2.8\textdegree{}C and mean risk ratio in excess of 26.
Averaging over the entire spatial domain, we find that under Mod4, the mean difference between the 100-year return levels based on 2018 and 1950 climates is
2.075\textdegree{}C, with 95\% credible interval of $[2.05, 2.10]$.  For Mod2 the mean difference is 2.320\textdegree{}C, with 95\% credible interval of $[2.30, 2.34].$  Similarly, under Mod4, the mean risk ratio over the entire spatial domain is 17.09, with 95\% credible interval $[16.86, 17.33]$,  so that on average a 100-year return level in the 1950 climate corresponds approximately to a 6-year return level in the 2018 climate.  
For Mod2, the mean risk ratio is increased to 18.77 with 95\% credible interval of $[18.57, 18.96]$. 

\begin{figure}[h]
\begin{center}
\includegraphics[scale=0.6]{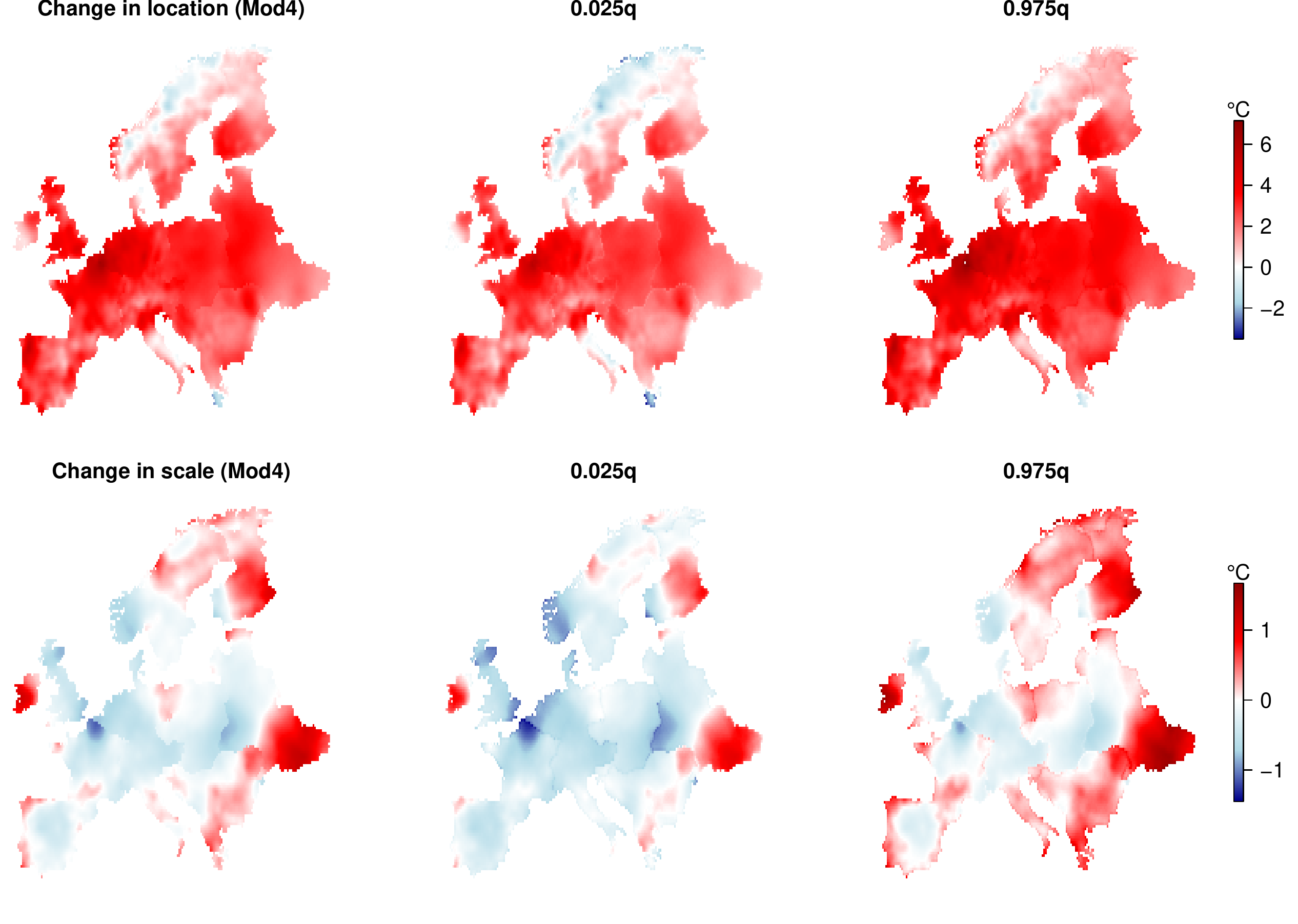} 
\caption{Changes in the GEV location and scale parameters over the period 1950-2018 (2018 parameter values subtract 1950 values) 
and  approximate 95\% credible interval limits, calculated by Monte Carlo simulation as described in Section \ref{Sec_RL_RR}, for Mod4.
Mod4 includes a trend in the GEV location and log-scale parameters using the covariate log($\textnormal{CO}_{2,t}/280$).}
\label{FittedSlopes}
\end{center}
\end{figure}

\begin{figure}
\begin{center}
  \includegraphics[scale=0.6]{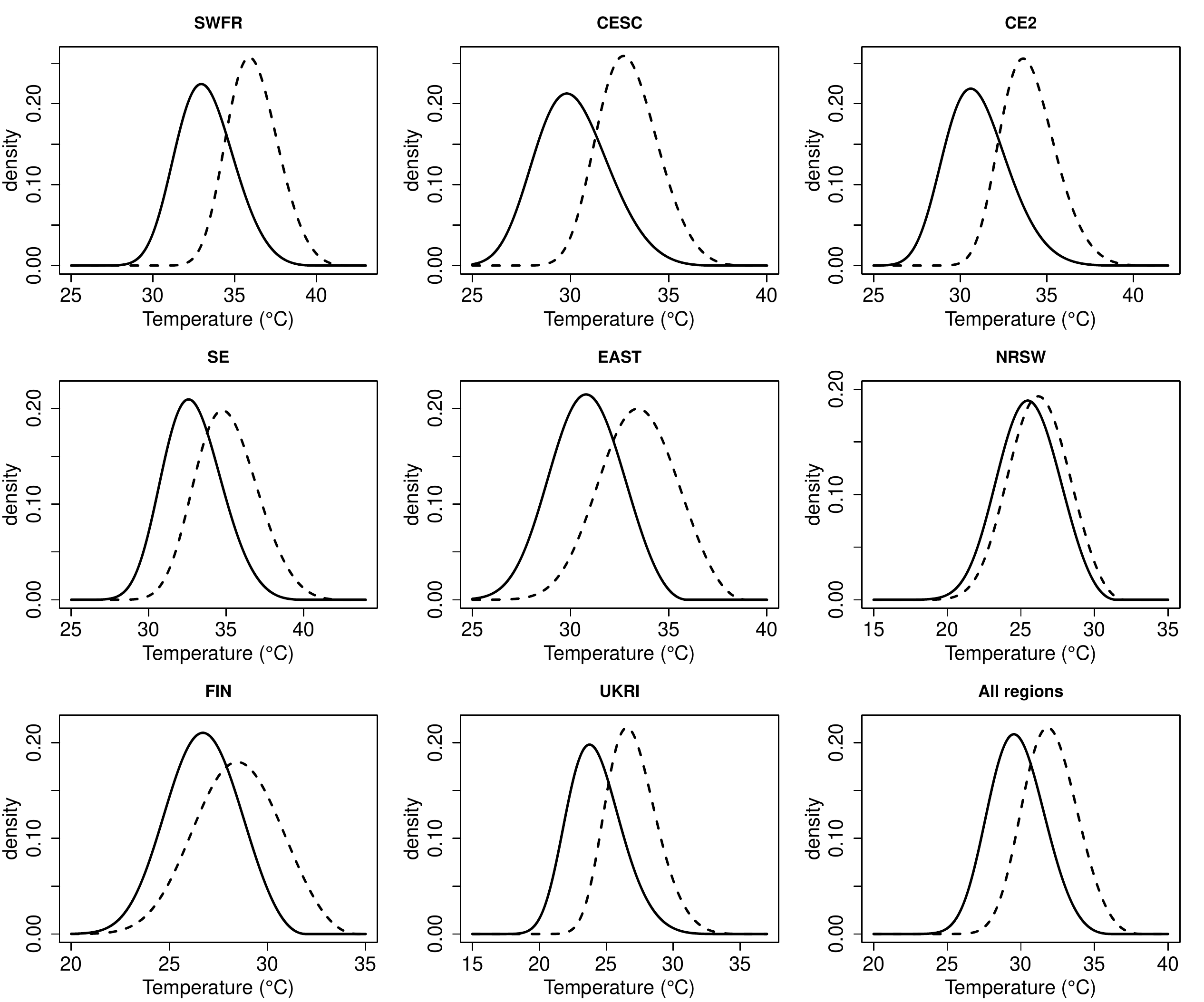}  
\caption{Plots showing the distribution (density function) of TXx based on spatially averaged values of the fitted GEV parameters in 1950
 (solid black curve) and 2018 (dashed curve), for each region defined in Table \ref{RegionSplit} as well as all regions together (bottom right plot). 
The GEV parameters in 1950 and 2018 are calculated using Mod4 which has trends in GEV location and log-scale parameters using the covariate log($\textnormal{CO}_{2,t}/280$).}
  \label{DensityChanges}
\end{center}
\end{figure}

Although of less interest to us than the return level differences and risk ratios, the changes in the GEV location and scale parameters over the period
1950 to 2018 calculated using Mod4 are shown in Figure \ref{FittedSlopes}.   
Approximately 95\% of the grid boxes show an increase in the location parameter over the study period, and after calculating approximate 95\% credible intervals for each change in location, 92\% of these have a lower limit greater than zero. We cannot however conclude that warming is detected in TXx in 
92\% of grid boxes as we make no attempt to correct for multiple comparisons. The changes in the GEV scale show rather more variability with $39\%$ positive and $61\%$ negative fitted values.  After calculating approximate 95\% credible intervals for the scale changes, 29\% of these slopes contain 0.  
The spatially averaged changes in the GEV location and scale parameters for each subregion in Table \ref{RegionSplit} are shown in
Table \ref{MeanChange} and the spatially averaged distributions of TXx in 1950 and 2018 are shown for each subregion in Figure \ref{DensityChanges}.
All regions show, on average, significant increases in the GEV location parameter, indicating a tendency for TXx to shift towards hotter temperatures.
There is a mixture of both increases and decreases in the spatially averaged GEV scale parameter differences corresponding to increasing and decreasing variability of TXx respectively. The regions with an average increase in the GEV scale parameter occur in the East as well as Ireland.
The changes in the GEV location parameter for Mod2 are the same as the return level differences shown in the bottom row of Figure \ref{RLDiffRas}.

In the same notation as Section \ref{StatMods}, for Mod4,  a test of the null hypothesis $\bm{\mu}_1 = \bm{0}$, i.e., all location slopes are zero, using the asymptotic distributional results of \cite{wood16}, has for each region a p-value less than $2\times 10^{-16}$. The same result is found for the location slopes from Mod2 and log-scale slopes for Mod4.  Together with the results in Table \ref{ModScores},
the hypothesis of no temporal variation in the GEV parameters is strongly rejected in all regions.

\clearpage

\section{Summary}  \label{DiscSec}

We have considered the problem of detecting and quantifying large scale changes in the distributions of the annual maximum daily maximum temperature (TXx) in a large subset of Europe during the years 1950-2018.  
Our approach was to divide the full domain in to 8 subregions over which several statistical models were fit.  In each of the models considered, TXx at each grid box was modelled using a generalized extreme value (GEV) distribution with the GEV location and scale parameters allowed to vary in time using atmospheric $\textnormal{CO}_2$ as a covariate.  We modelled the GEV parameters 
as varying smoothly over space, where the appropriate degree of smoothness was determined objectively using the methods of \cite{wood16}. 
Changes were detected most strongly in the GEV location parameter with the distributions of TXx shifting towards hotter temperatures at most grid boxes.
Although the best performing model in all regions has both the GEV location and scale parameters changing in time, the signal for changes in the scale parameters is noisier than that for the location parameters with some regions showing a tendency for increases in scale and others for a decrease.
The regions that show tendency for an increase in scale, corresponding to an increase in the variability in TXx, are in Eastern Europe and Ireland.
The second best performing model in all regions has only the GEV location parameter changing in time. 
Regardless of whether our best or second best models were used, our main findings regarding changes in return levels based on the 2018 and 1950 climates as well as risk ratios broadly agree.  Using our best performing model and averaging across our entire spatial domain, the 100-year return level of TXx based on the 2018 climate is approximately 2\textdegree{}C hotter than that based on the 1950 climate. Also averaging across our spatial domain, the 100-year return level of TXx based on the 1950 climate corresponds approximately to a 6-year return level in the 2018 climate.  Our findings are most robust in Central Europe where the underlying network of weather stations used to construct the gridded dataset has the highest density.

\vspace{20mm}

\textbf{Acknowledgement:} We are grateful to Ben Youngman for helpful advice regarding use of the \texttt{evgam} package.

\begin{table}  
\centering
\caption{Comparison of model scores defined in Appendix \ref{ModScoring} by regions as defined in Table \ref{RegionSplit}. The smallest, i.e., best, scores for each region are in bold.} 
 \label{ModScores}
\begin{tabular}{|c | c | c | c | c | c | c |}
 \hline 
	 \textbf{Region} &  \textbf{Model}  & \textbf{SE} & \textbf{DS} & \textbf{CRP} & \textbf{WCRP} &  \textbf{AIC} \\
\hline
 SWFR  &  Mod1 &  3.8357 &  2.2769 & 1.0868 &  0.3358 & 559915.2 \\
  &  Mod2  & 3.0011  &   2.0374  &  0.9582 &  0.3005  &  525032.2  \\
  &  Mod3  & 3.7451  &  2.2538  &  1.0722   &  0.3231  &   556665.9  \\
  &  Mod4  &  \textbf{2.9925} &  \textbf{2.0257} & \textbf{0.9557}  & \textbf{0.2995} &  \textbf{523671.7} \\
  &  Mod5  &  3.0612  & 2.0583 &  0.9696  &  0.3032 &  529017.3  \\
\hline
 CESC &  Mod1 &  3.9895  &  2.3430  & 1.1231 &  0.3470  &  455694.1  \\
  &  Mod2  &  \textbf{3.1275}  &  2.1097  &  0.9876 &  0.3090  & 429074.8 \\
  &  Mod3  &  3.8982  &  2.3157 &  1.1061  &  0.3341  &  452628.2 \\
  &  Mod4  &  3.1284  &  \textbf{2.0959} & \textbf{0.9868} &  \textbf{0.3081} &  \textbf{427694.2} \\
  &  Mod5  &  3.2606  &  2.1510  &  1.0101 & 0.3146 &  433790.1  \\
\hline
 CE2 & Mod1 &  3.9107 &  2.3593 &  1.1163 &  0.3504 &  323433.1 \\
        & Mod2  &  3.1112  &  2.1293 &  0.9865 &  0.3138  &  304560.4   \\
        & Mod3  &  3.8199  &  2.3348  & 1.1025   &  0.3396  &  322131   \\
        & Mod4  &  \textbf{3.1084}  &  \textbf{2.1196}  & \textbf{0.9858}   &  \textbf{0.3133}   &  \textbf{303760.9}   \\        
        & Mod5  &  3.1197  &  2.1321 & 0.9877  &  0.3142 &   304705.4   \\    
\hline
 SE & Mod1 &  4.3495  &  2.4603 & 1.1690 &  0.3660  &  390091.8    \\
      & Mod2 &  3.8561  &  2.3336 & 1.1028 &  0.3446 &  377784.1  \\ 
      & Mod3 &  4.2567  &  2.4290   &  1.1555  & 0.3542  &  387405.2   \\ 
      & Mod4 &  \textbf{3.8560}  & \textbf{2.3299} &  \textbf{1.1025} & \textbf{0.3445} &  \textbf{377420.2} \\ 
      & Mod5 &  3.9081 & 2.3470 &  1.1109 &  0.3473  &  379074.1   \\ 
\hline
 EAST & Mod1 & 4.0814 & 2.4005  &  1.1429  & 0.3521 &  568594.9 \\
          & Mod2 & 3.4370 & 2.2228  &  1.0502  & 0.3189 &  544041.1  \\
          & Mod3 & 3.9377 & 2.3404  &  1.1142  & 0.3305   &  557883.2 \\
          & Mod4 & \textbf{3.4310}  & \textbf{2.2140} &   \textbf{1.0475} &  \textbf{0.3179}  &  \textbf{542028.1}  \\
          & Mod5 & 3.4499 & 2.2269 &  1.0527 &  0.3195  &  544577.9  \\
\hline
  NRSW & Mod1  & 4.3202  & 2.4486  & 1.1719 & 0.3555 &  620146  \\
            & Mod2  & 4.1793  & 2.4148  & 1.1525 & 0.3492 & 615169.8   \\
            & Mod3  & 4.3163  &  2.4436 & 1.1707 & 0.3546 &  619132.8  \\ 
            & Mod4  & \textbf{4.1789}  &  \textbf{2.4111} & \textbf{1.1520} & \textbf{0.3491} &  \textbf{614354.3}   \\ 
            & Mod5  & 4.2745 &  2.4369 & 1.1664 & 0.3532 &  618132.7   \\ 
\hline
  FIN & Mod1 &  4.2185  & 2.4237 & 1.1524  &  0.3488  &  287681.3  \\
        & Mod2 &  3.8567 &  2.3377 &  1.1060  &  0.3337  &  281865.8  \\
        & Mod3 &  4.1635  &  2.4116  &  1.1435  & 0.3404  &  286201.7  \\
        & Mod4 &  \textbf{3.8564}  &  \textbf{2.3226}  &  \textbf{1.1035} &  \textbf{0.3331} & \textbf{280744}  \\
        & Mod5 &  3.9453  &  2.3589  & 1.1175 & 0.3353  &  283438.9  \\
\hline
UKRI & Mod1 & 4.8823 &  2.5763 & 1.2464  &  0.3862   & 186346.2   \\
        & Mod2 & 4.0867 &  2.3989 &  1.1400 &  0.3599   & 178130.3    \\
        & Mod3 & 4.8413 &  2.5652  & 1.2399  &  0.3816  & 185814.7      \\
        & Mod4 &  \textbf{4.0813} & \textbf{2.3827}  &  \textbf{1.1353}  &  \textbf{0.3580} & \textbf{177302.5} \\
        & Mod5 &  4.1584 & 2.4141 & 1.1479 &  0.3595  & 179283  \\
\hline
\end{tabular}
\label{ModScores}
\end{table}

\begin{table}  
\centering
\caption{Approximate 95\% credible intervals for the spatially averaged 100-year return level differences, in \textdegree{}C, based on 2018 and 1950 climates (2018 return level subtract 1950 return level) and risk ratios for each region defined in Table \ref{RegionSplit}. The results are based on Mod4, which includes a trend in the GEV location and log-scale parameters using the covariate log($\textnormal{CO}_{2,t}/280$). The endpoints of the intervals were calculated using Monte Carlo simulation.  We simulated values of the GEV parameters from their posterior distributions at each grid box based on the 2018 and 1950 climates.  We then calculated the return level differences and risk ratios at each grid box and calculated the mean across the region. This procedure was repeated 2000 times.  The $\alpha/(2\times\,8)$ and $1 - \alpha/(2\times\,8)$, with $\alpha = 0.05$, empirical quantiles of the 2000 estimated means give the left and right endpoints respectively of the intervals shown, where we have corrected for multiple comparisons using the Bonferroni correction.}  
 \label{ModScores}
\begin{tabular}{|c | c | c |}
 \hline 
	 \textbf{Region} &  \textbf{Spatially averaged}  & \textbf{Spatially averaged}  \\
                              &             \textbf{return level difference}  & \textbf{risk ratio}    \\
\hline
         SWFR       &   [2.25, 2.41]      &  [20.57, 22.36]   \\
         CESC        &    [1.88, 2.08]     &    [17.63, 19.48]  \\
         CE2          &    [2.15, 2.39]     &    [14.73, 17.21]  \\ 
         SE            &    [2.32, 2.56]     &    [15.22, 16.85]  \\
        EAST         &   [2.72, 2.85]      &    [25.61, 27.31]  \\
        NRSW        &   [0.51, 0.66]      &    [4.20, 4.91]    \\
        FIN            &   [2.39, 2.59]      &    [18.81, 20.73] \\
        UKRI          &    [2.25, 2.69]     &   [11.48, 14.10]  \\ 
\hline
\end{tabular}
\label{MeanDiff} 
\end{table}

\begin{table}  
\centering
\caption{P-values from test of pairwise exchangeability of CRP/WCRP scores for the two best models, Mod4 and Mod2, by region.} 
 \label{PVals}
\begin{tabular}{|c | c | c |}
 \hline 
	 \textbf{Region} & \textbf{CRP p-value}  & \textbf{WCRP p-value}  \\
\hline
        SWFR    &  $< 10^{-6}$  &  $< 10^{-6}$   \\
       CESC      &   $< 10^{-6}$ &  $< 10^{-6}$  \\
        CE2         &   $< 10^{-6}$   &  $< 10^{-6}$     \\
        SE            &  $3 \times 10^{-6}$    &   $8.4 \times 10^{-5}$   \\
        EAST        &   $< 10^{-6}$   &   $ < 10^{-6} $   \\   
         NRSW       &  $< 10^{-6}$   &   $4.3 \times 10^{-5}$  \\
        FIN            & $< 10^{-6}$    &   $ < 10^{-6} $   \\
        UKRI          &   $< 10^{-6}$  &   $ < 10^{-6} $    \\ 
\hline
\end{tabular}
\label{PVals} 
\end{table}

\begin{table}  
\centering
\caption{Approximate 95\% credible intervals for the spatially averaged changes in the GEV location and scale parameters (2018 parameter value subtract 1950 parameter value) for each region defined in Table \ref{RegionSplit}.  Calculations are based on Mod4 which has trends in GEV location and log-scale parameters using the covariate log($\textnormal{CO}_{2,t}/280$). The endpoints of the intervals are calculated by Monte Carlo simulation as described in the caption to Table \ref{MeanDiff}. }  
 \label{AverageParamChanges}
\begin{tabular}{|c | c | c |}
 \hline 
	 &  \textbf{Spatially averaged}  & \textbf{Spatially averaged}  \\
         \textbf{Region}                      &             \textbf{change in}  & \textbf{change in}    \\
                 &             \textbf{location}  & \textbf{scale}    \\
\hline
         SWFR       &   [2.95, 3.02]     &  [-0.24, -0.19]   \\
         CESC        &    [2.93, 3.00]     &    [-0.34, -0.29]     \\
         CE2          &    [3.03, 3.12]     &    [-0.28, -0.22]   \\ 
         SE            &    [2.11, 2.21]     &    [0.07, 0.14]   \\
        EAST         &   [2.54, 2.65]      &    [0.10, 0.17]    \\
        NRSW        &   [0.69, 0.77]      &    [-0.07, -0.02]    \\
        FIN            &   [1.58, 1.70]      &    [0.28, 0.35]   \\
        UKRI          &    [2.76, 2.89]     &   [-0.20, -0.10]    \\ 
\hline
\end{tabular}
\label{MeanChange} 
\end{table}

\clearpage

\appendix

\section{Model scoring} \label{ModScoring}

While we report model AIC values and perform hypothesis tests, we also consider another method of model comparison that does not depend on asymptotic theory.  We use several scoring rules that evaluate the performance of models based on their ability to predict unseen data, i.e., data that was held out from the model fitting procedure.  A scoring rule is a function, $S$, that assigns a real number value $S(F,y)$, to the pair $(F,y)$ where $F$ is a predictive distribution and $y$ is an observed value.  In our case $F$ will be the cumulative distribution function of a fitted GEV distribution and $y$ will be some observation, held out from the model fitting procedure, that under our model is assumed to be drawn from $F$.  

All of the scoring rules that we consider are negatively oriented so that smaller scores correspond to better predictions.
A negatively oriented scoring rule $S$ is called proper \citep{gneit07} if 
\begin{equation}
\mathbb{E}\{S(G, Y)\} \leq \mathbb{E}\{S(F, Y)\} \quad \text{when}\, Y\sim G,  \label{ProperDefn}
\end{equation}
where $\mathbb{E}$ denotes expectation, i.e., on average, the true distribution $G$ will not give a worse score than any other distribution $F$, so that forecasters are incentivized to report the truth.
All the scores that we consider are proper.  If equality in (\ref{ProperDefn}) holds only if $F=G$ then $S$ is called strictly proper.  

One of the simplest, and most common scoring rules is the squared error score, $S_{SE}$ defined by
\begin{equation}
S_{SE}(F,y) = (y - \mu_F)^2
\end{equation}
where $\mu_F$ is the expected value of a random variable with distribution $F$.  The squared error score gives higher penalties the further an observation is from the 
mean of the predictive distribution, but may be regarded as rather simplistic in that the whole predictive distribution $F$ is represented in the score only through it's mean 
$\mu_F$.  When the variance of the predictive distribution $F$ is large, we may wish to give a smaller penalty to observations $y$ that are far from the mean $\mu_F$.
One score that does this is the Dawid-Sebastiani score, $S_{DS}$, defined by 
\begin{equation}
S_{DS}(F, y)  = \bigg(\frac{y - \mu_F}{\sigma_F}\bigg)^2 + \text{log}\,\sigma^2_F
\end{equation}
where $\sigma^2_F$ is the variance of $F$. 

Another commonly used scoring rule is the continuous ranked probability score (CRPS) defined by 
\begin{equation}
S_{CRP}(F,y) = \int_{-\infty}^{\infty}\big(F(x) - \mathbbm{1}[y\leq x]\big)^2dx.  \label{CRPS1}
\end{equation}
Unlike the squared error and Dawid-Sebastiani scores, the CRPS takes into account the full predictive distribution and compares it with the empirical distribution based on the single observation $y$.  A closed form expression for the CRPS when $F$ is the distribution function of a GEV random variable is given in \cite{fried12} and implemented in the R package \texttt{scoringRules} \citep{scoringRules}. 

For an extreme value analysis, it may be desirable to consider a scoring rule that gives a higher penalty for poor prediction in the tails of the distribution.
One way to do this is to use a weighted version of the CRPS \citep{gneit11}.  First we note an alternative representation of the CRPS in terms of quantiles is
\begin{equation}
S_{CRP}(F,y) = 2\int_0^1 \big(\mathbbm{1}[y\leq F^{-1}(p)] - p\big)\big(F^{-1}(p) - y \big)dp.  \label{CRPS2}
\end{equation}
The equality of (\ref{CRPS1}) and (\ref{CRPS2}) is shown in \cite{laio08}.  
The weighted continuous ranked probability score (WCRPS) is obtained from (\ref{CRPS2}) by adding an extra factor, $w(p)$, to the integrand, which determines the weight
given to the $p$-th quantile giving 
\begin{equation}
S_{WCRP}(F, Y) = 2\int_0^1 \big(\mathbbm{1}[y\leq F^{-1}(p)] - p\big)\big(F^{-1}(p) - y \big)w(p)dp.  \label{WCRPS}
\end{equation}
In our application we use the weighting function $w(p) = p^2$.  
As the integral (\ref{WCRPS}) does not have a closed form solution when $F$ is the distribution function of a GEV random variable, we approximate it with the summation
\begin{equation}
\hat{S}_{WCRP}(F, Y) = \frac{2}{N} \sum_{i=1}^N \big(\mathbbm{1}[y\leq F^{-1}(p_i)] - p_i\big)\big(F^{-1}(p_i) - y \big)w(p_i)
\end{equation}
for large $N$ and $0 \leq p_1 < p_2 < \ldots < p_n \leq 1$ a partition of the interval $[0,1]$.  In our case we take the evenly spaced partition $p_i = i/N$ 
with $N = 1000$ and $i =1,2,3,\ldots,999$.

We use each of the scoring rules described above as part of a cross-validation scheme to evaluate a models performance.  This is described in
Appendix \ref{CV_Sec}.

\section{Cross-validation and score comparisons} \label{CV_Sec}

For each of the regions in Table \ref{RegionSplit}, we evaluate the performance of each of the models in Table \ref{ModTrends} using 5-fold cross-validation.
Specifically, for a fixed region, we randomly assign each observation of the region to one of five subsets, or splits, of approximately equal size.  The same splits are used in each model evaluation. 
Suppose there are $N_j$ observations in split $j, 1\leq j \leq 5$, which we denote by $y^{(j)}_1, y^{(j)}_2,\ldots, y^{(j)}_{N_j}$. 
We fix one of the splits, $k$ say, $1\leq k \leq 5$, to be used as test data and fit the model of interest to the data from the remaining four splits. 
For each observation $y^{(k)}_i, 1\leq i \leq N_k$, from the test data, we evaluate  $S(F^{(k)}_i, y^{(k)}_i)$ for each of the scores defined in Appendix \ref{ModScoring}
where $F^{(k)}_i$ is the GEV distribution function that under the fitted model, $y^{(k)}_i$ is assumed to be drawn from.
This procedure is carried out for each $k, 1\leq k \leq 5$, so that each split gets used as test data.  
The mean score, $\frac{1}{N}\sum_{k=1}^{5}\sum_{i=1}^{N_k}S(F^{(k)}_i, y^{(k)}_i)$, where $N= \sum_{k=1}^{5}N_k$ is the total number of observations in the region, gives an overall measure of model performance according to score $S$.  For the scores defined in Appendix \ref{ModScoring}, which are all negatively oriented, models with lower mean score are preferred. 

Suppose that in the cross-validation procedure described above, model $B$, produces a lower mean score than another model, $A$. If the observed difference in the mean scores 
is very small, we may wish to test whether this really provides evidence that model $B$ is better than $A$.  Suppose for each model, we have the $N$ scores
$S(F_i^A, y_i)$ and $S(F_i^B, y_i)$, $1\leq i \leq N$, where $F_i^A$ and $F_i^B$ are the distribution functions that under models $A$ and $B$ respectively, observation $y_i$ is assumed to be drawn from. We will construct a test for the null hypothesis that the scores $S(F_i^A, y_i)$ and $S(F_i^B, y_i)$ are pairwise exchangeable, for $1\leq i \leq N$. 
Two random variables $X_1$ and $X_2$ are said to be exchangeable if $\mathbb{P}(X_1 \leq x_1, X_2 \leq x_2) = \mathbb{P}(X_1 \leq x_2, X_2 \leq x_1).$
In particular, this implies that $X_1$ and $X_2$ are identically distributed.  
If the scores $S(F_i^A, y_i)$ and $S(F_i^B, y_i)$ are pairwise exchangeable, then the score $S(F_i^A, y_i)$ would be equally likely to have been produced by model $B$ and 
similarly $S(F_i^B, y_i)$ equally likely to have been produced by model $A$. Thus for the observed score difference $S_i^- = S(F_i^A, y_i) - S(F_i^B, y_i)$, we would have been equally likely to have observed $-S_i^-$ under the null hypothesis. This motivates the following test procedure defined in Algorithm \ref{TestAlg} below. 

\begin{algorithm}
\SetKwData{Left}{left}\SetKwData{This}{this}\SetKwData{Up}{up}
\SetKwFunction{Union}{Union}\SetKwFunction{FindCompress}{FindCompress}
\SetKwInOut{Input}{input}\SetKwInOut{Output}{output}

\Input{Positive integer $J$ and model scores $S(F_i^A, y_i), S(F_i^B, y_i)$, $1 \leq i \leq N$.}
\Output{$p$, an estimate of the p-value of the test.}
\For{$i\leftarrow 1$ \KwTo $N$}{
\emph{compute the score difference}\;
          $S_i^- = S(F_i^A, y_i) - S(F_i^B, y_i) $ \;
}
\emph{compute the observed test statistic $T_{obs} = \frac{1}{N}\sum_{i=1}^{N}S_i^- $} \;
\For{$j\leftarrow 1$ \KwTo $J$}{
       \For{$i\leftarrow 1$ \KwTo $N$}{  
                                                      \emph{compute the randomized score difference}\;
                                                                           \begin{equation*}
                                                            S_i^-(j) =   \begin{cases}
                                                                 S_i^-  \,\,\, & \text{with probability $0.5$}  \\
                                                                 -S_i^-  \,\,\, & \text{with probability $0.5$}  \\
                                                                                \end{cases}
                                                                              \end{equation*}
                                                        }
\emph{compute} $T_j =  \frac{1}{N}\sum_{i=1}^{N}S_i^-(j) $  \;
}
\emph{return} $p =  \frac{1}{J}\sum_{j=1}^{J} \mathbbm{1}[T_j \geq T_{obs}].$
\caption{Hypothesis testing procedure for testing pairwise exchangeability of model scores.}\label{TestAlg}
\end{algorithm}
The value of $p$ computed in  Algorithm \ref{TestAlg} is an unbiased estimate of the one sided  p-value for the test with null hypothesis that the scores $S(F_i^A, y_i)$ and $S(F_i^B, y_i)$ are pairwise exchangeable, $1 \leq i \leq N$.  The value of the observed test statistic $T_{obs}$ is strictly positive since, it is assumed that model $B$ has the lower observed mean score.  Small values of $p$ give evidence against the null hypothesis in favour of model $B$.  In our applications of Algorithm 
 \ref{TestAlg} we take $J=10^6$.

\section{Diagnostic plots} \label{App1}

In this Appendix we perform some visual checks for Mod4, to see whether this model provides a reasonable fit to the data and is not merely the best of a bad bunch of models.
As it is not feasible to provide plots for every grid box, we consider the performance as a whole over the subregions as defined in Table \ref{RegionSplit}.

\begin{figure}[h]
\begin{center}
\includegraphics[scale=0.5]{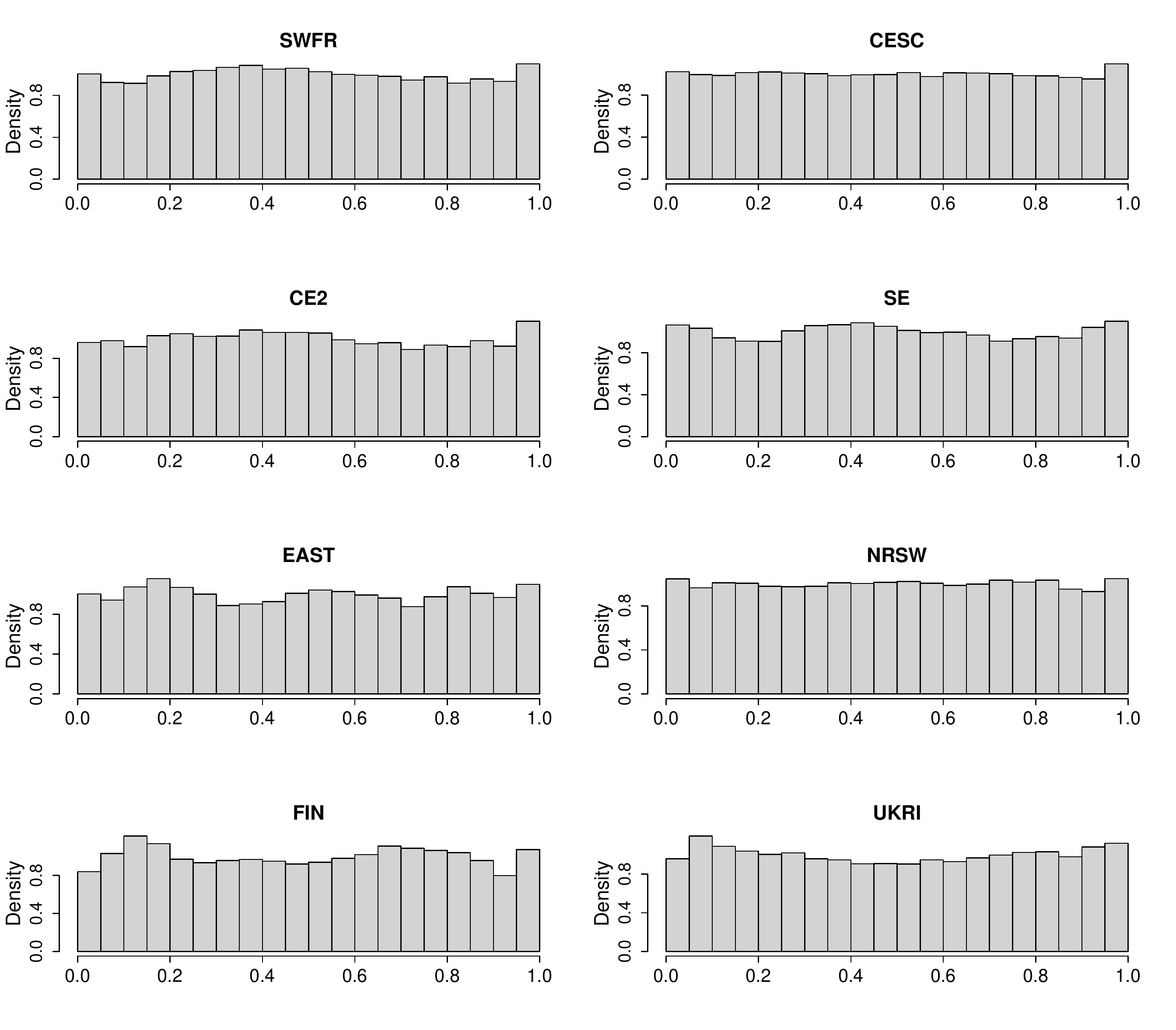} 
\caption{Histograms of probability integral transform values by region for Mod4.}
\label{PITMod4}
\end{center}
\end{figure}

One simple way to check for any systematic discrepancies between a fitted model and the observed data is to use the probability integral transform (PIT). 
The PIT states that if $Y$ is a random variable with continuous distribution function $F$ then the random variable $F(Y)$ is uniformly distributed between 0 and 1.
Suppose that, given a sample of size $n$ with observed response variables, $y_i$, $1\leq i \leq n$, a statistical model fits distribution function $F_i$ to $y_i$.
Then if the model is correct the values $F_i(y_i)$, $1\leq i \leq n$, are a sample of size $n$ from a uniform distribution on $[0, 1]$. The plausibility of this may be checked visually, 
e.g., by plotting a histogram of the $n$ PIT values $F_i(y_i)$, $1\leq i \leq n$.  A U-shaped histogram would indicate that the fitted model is underdispersive, i.e., it is not
adequately accounting for the variability in the data, whereas a histogram that is too peaked in the middle indicates the model is overdispersive.
Histograms of the PIT values, by region, are shown for Mod4 in Figure \ref{PITMod4} and do not show any serious cause for concern.

\begin{figure}[h]
\begin{center}
\includegraphics[scale=0.5]{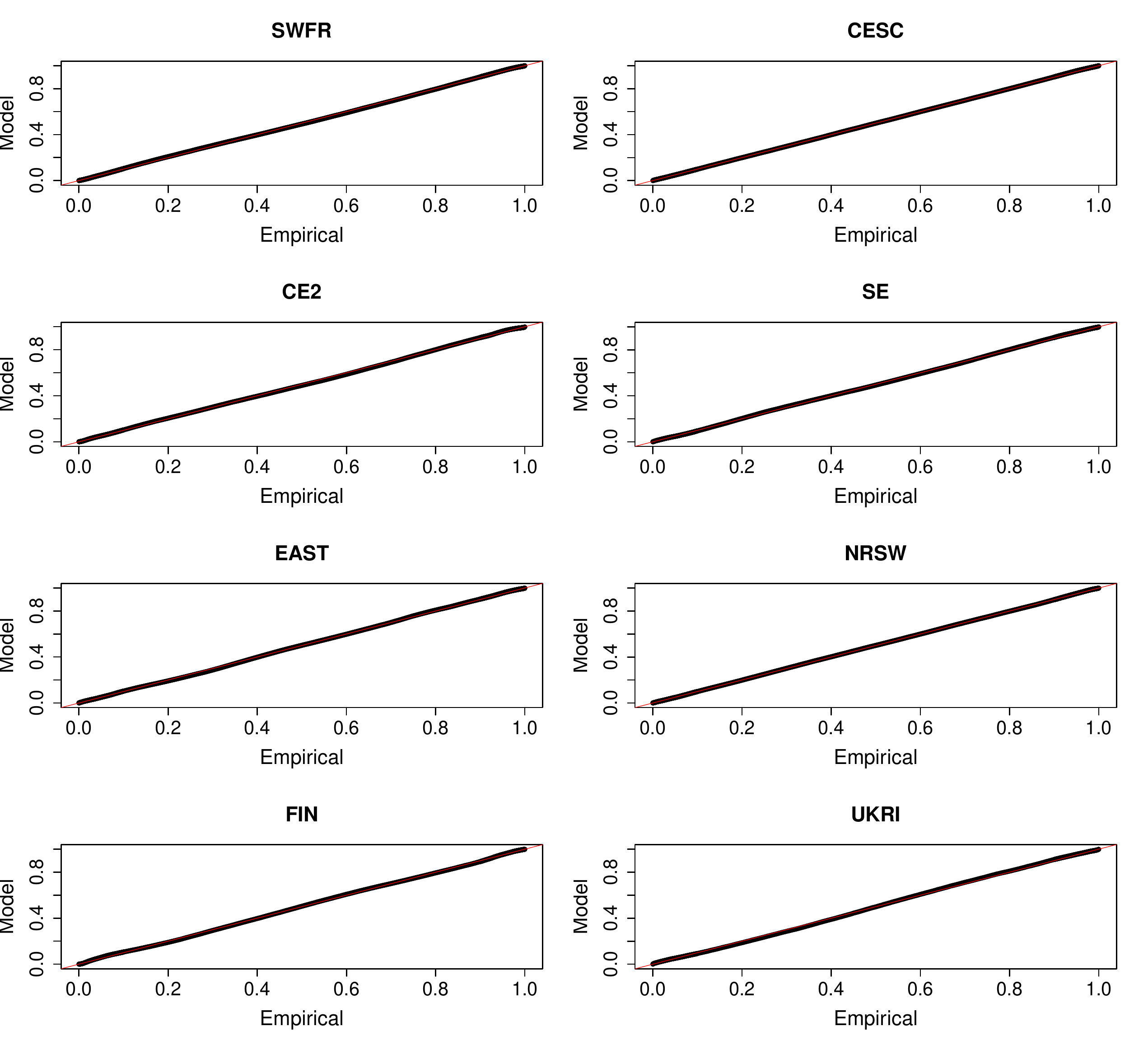} 
\caption{Probability plots by region for Mod4.}
\label{pp_Mod4}
\end{center}
\end{figure}

\begin{figure}[h]
\begin{center}
\includegraphics[scale=0.5]{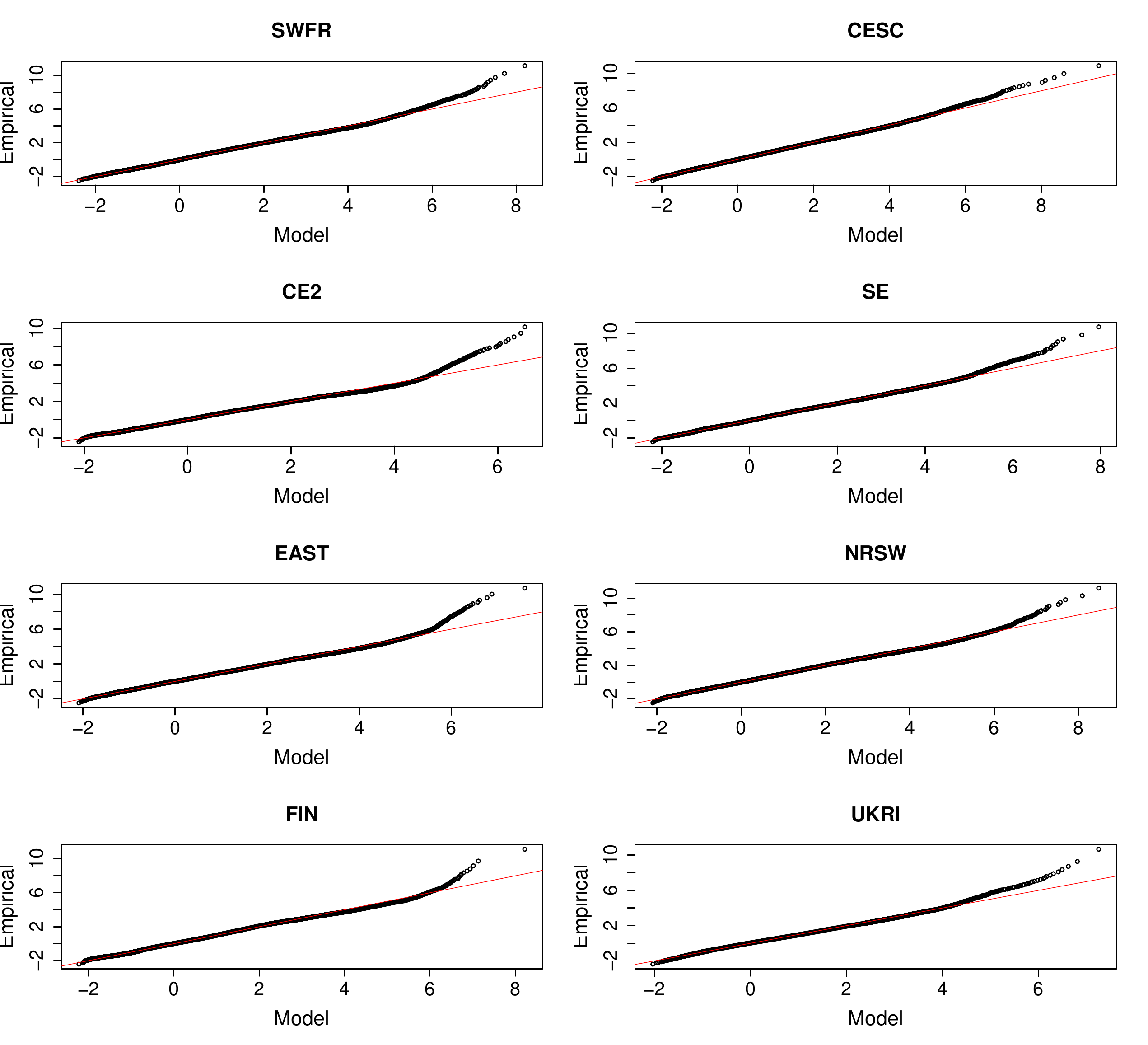} 
\caption{Quantile-quantile plots by region for Mod4.}
\label{qq_Mod4}
\end{center}
\end{figure}

Another standard method for checking a non-stationary extreme value model fit is via probability or quantile plots \citep[Section 6.2.3]{cole01}.
Both of these plots are based on the fact that if $Y_{it}$ is a GEV random variable with parameters $\bm{\Psi}_{it}= (\mu_{it}, \sigma_{it}, \xi_{it})$ then the variable
$Z_{it}$ defined by
\begin{equation}
Z_{it} = \frac{1}{\xi_{it}}\text{log}\bigg\{1 + \xi_{it}\bigg(\frac{Y_{it} - \mu_{it}}{\sigma_{it}}\bigg)\bigg\}  \label{Z_Defn}
\end{equation}
has a standard Gumbel distribution with distribution function $F(z) = \text{exp}\{-\text{exp}(-z)\}, z\in \mathbb{R}$.
If we fix a specific region in Table \ref{RegionSplit}, and suppose that as in Section \ref{GEV_Sec}, in grid box $i$ we have observed annual maxima,
$y_{it_j}, 1\leq j \leq n_i$,  then from a given fitted model, we may obtain the values $z_{it_j}, 1\leq j \leq n_i$, by applying the transformation 
(\ref{Z_Defn}).  If there are $N$ grid boxes in the region in total then we may apply this transformation to all grid boxes and obtain 
$m = \sum_{i=1}^{N}n_i$ transformed $z$ values.  If the model were correct, then the ordered values $z_{(1)}, z_{(2)}, \ldots z_{(m)}$ where
$z_{(k)} \leq z_{(l)}$ when $k\leq l$, would be a sample from 
a standard Gumbel distribution.  The probability plot tests the plausibility of this by comparing empirical and fitted model probabilities and plots the pairs
\begin{equation*}
 \bigg(  \frac{k}{m+1}, \text{exp}\{-\text{exp}(-z_{(k)})  \}   \bigg), \quad  1\leq k \leq m.
\end{equation*}
The quantile plot compares fitted model and empirical quantiles and plots the pairs
\begin{equation*}
 \bigg(z_{(k)},  -\text{log}\bigg\{-\text{log}\bigg(\frac{k}{m+1}\bigg)\bigg\}  \bigg), \quad  1\leq k \leq m.
\end{equation*}
The further these plots deviate from a diagonal line, the greater the discrepancy between the fitted model and the observations.  Probability and quantile plots are shown by region 
in Figures \ref{pp_Mod4} and \ref{qq_Mod4}.  The probability plots stay very close to the diagonal line indicating a good quality of fit for each subregion. The quantile plots show  to a varying extent in each region, some discrepancy in the upper tail.  For sake of reference, the values 5,6,7 and 8 correspond to the 0.9933, 0.9975, 0.9991 and 0.9997 theoretical quantiles of the standard Gumbel distribution respectively. Thus we can see that Mod4 is giving a good fit in all regions at least up to the 0.9933 regional quantile and in several regions beyond this
but fails to explain approximately the largest 0.05\% of observations in each region.

\begin{figure}
  \centering
\begin{subfigure}{0.9\textwidth}
  \centering
\includegraphics[scale=0.5]{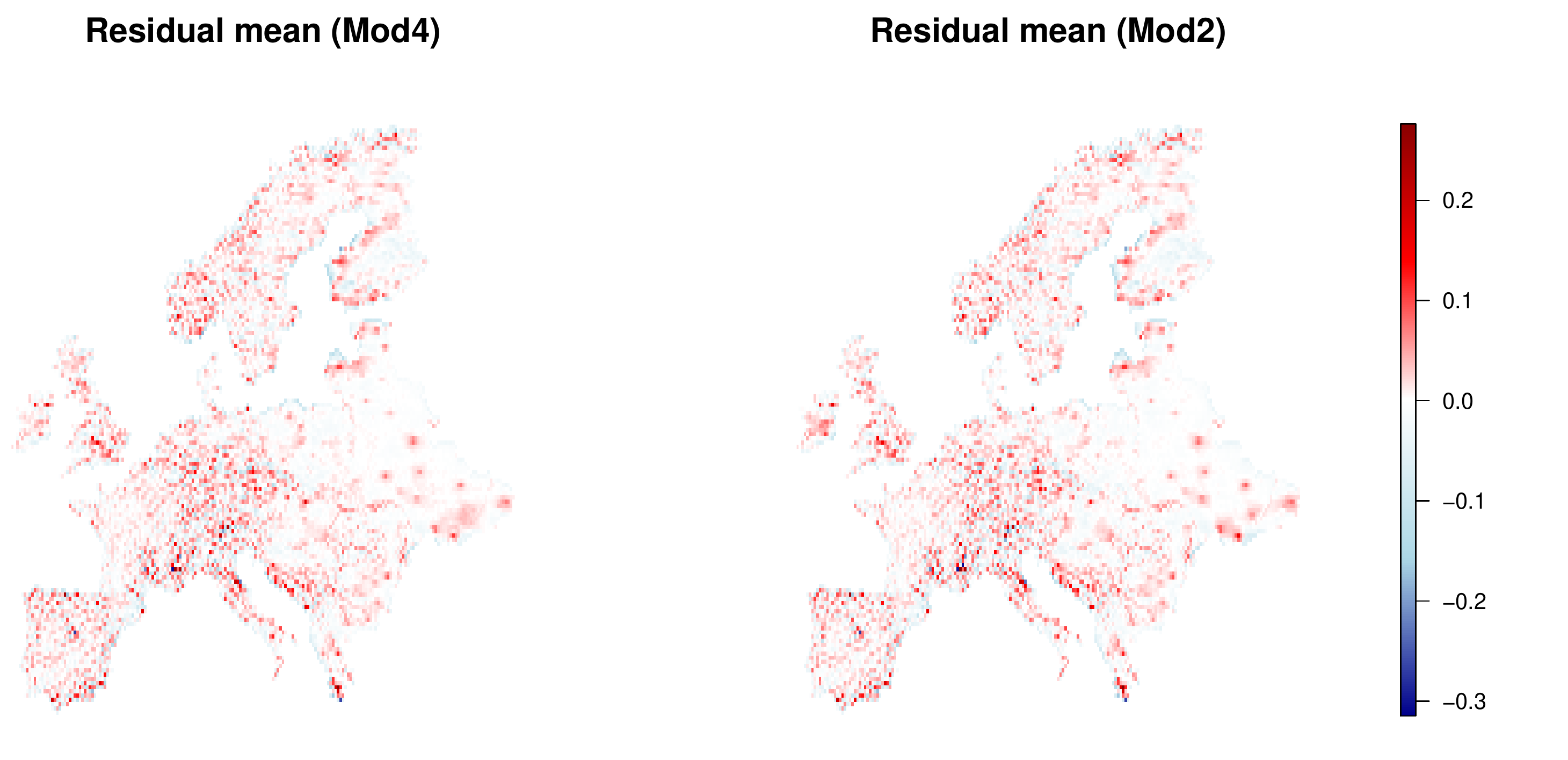}   
  \label{PRmean}
\end{subfigure}

\begin{subfigure}{0.9\textwidth}
  \centering
\includegraphics[scale=0.5]{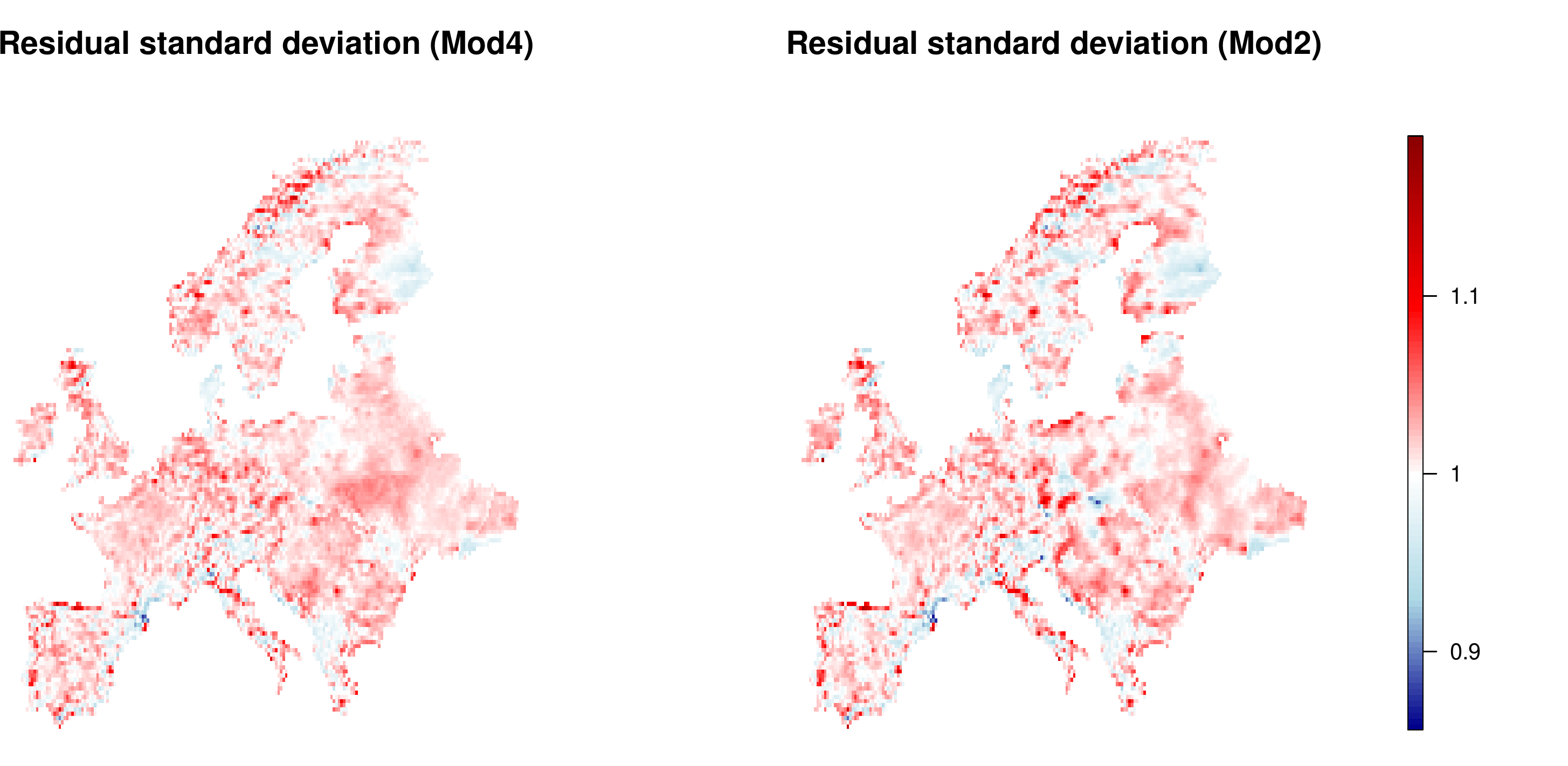} 
  \label{PRsd}
\end{subfigure}
\caption{Mean values (top row) and standard deviations (bottom row) of Pearson residuals for Mod4 and Mod2.}
\label{PearsonRes}
\end{figure}

Finally, we inspect the spatial distribution of the Pearson residuals obtained from a model fit.  For grid box $i$ we obtain the $n_i$ residuals 
$r_{ij}, 1\leq j \leq n_i$ defined by
\begin{equation}
r_{ij} = \frac{y_{it_j} - \hat{\mathbb{E}}(y_{it_j})}{\{\widehat{\text{Var}}(y_{it_j})\}^{1/2}} 
\end{equation}
where $\hat{\mathbb{E}}(y_{it_j})$ and $\{\widehat{\text{Var}}(y_{it_j})\}^{1/2}$ denote the model fitted expected value and standard deviation respectively.
When the fitted distribution is GEV with parameter vector $\bm{\Psi}_{it}= (\mu_{it}, \sigma_{it}, \xi_{it})$, then these expressions become
\begin{align}
\hat{\mathbb{E}}(y_{it_j}) & =  
 \begin{cases}
 \mu_{it} + \frac{\sigma_{it}}{\xi_{it}}\{\Gamma\,(1 - \xi_{it}) - 1\}, \quad   & 0 < \xi_{it} < 1, \\
  \mu_{it} +    \sigma_{it}\,\gamma,          \quad  & \xi_{it} = 0,  \\
  \infty,                                                     \quad  & \xi_{it} \geq 1  \\
\end{cases} 
\intertext{and}
\widehat{\text{Var}}(y_{it_j}) & = 
 \begin{cases}
   \frac{\sigma^2_{it}}{\xi^2_{it}} \{\Gamma\,(1-2\xi_{it}) - \Gamma^2(1 - \xi_{it})\}  ,\quad   & 0 < \xi_{it} < 1/2, \\[10pt]
  \frac{\pi^2\sigma_{it}^2}{6},    \quad  & \xi_{it} = 0,  \\[10pt]
  \infty,                          \quad  & \xi_{it} \geq 1/2,  \\
\end{cases} 
\end{align}
where $\gamma \approx 0.5772$ is Euler's constant and $\Gamma(t) = \int_{0}^{\infty}x^{t-1}e^{-x}dx.$
If the fitted model were correct then the Pearson residuals are realizations of a random variable with mean 0 and standard deviation 1.
The mean and standard deviation of the Pearson residuals for each grid box for both Mod4 and Mod2 are shown in Figure \ref{PearsonRes}. 
The spatial distributions look similar for both models and don't display any worryingly large deviations from the zero mean, unit standard deviation assumption. 

\end{document}